\documentclass[11pt,twoside,letterpaper]{article} %% The same for {book}
\usepackage{times,fancyhdr}
\usepackage{graphicx}


\makeatletter 
\def\cleardoublepage{\clearpage\if@twoside \ifodd\c@page\else% 
    \hbox{}% 
    \thispagestyle{empty}%
    \newpage% 
    \if@twocolumn\hbox{}\newpage\fi\fi\fi} 
\makeatother

\def\figurename{Figure}
\makeatletter
\renewcommand{\fnum@figure}[1]{\figurename~\thefigure.}
\makeatother

\def\tablename{Table}
\makeatletter
\renewcommand{\fnum@table}[1]{\tablename~\thetable.}
\makeatother

\begin{document}
\title{
{\begin{flushleft}
\vskip 0.45in
%{\normalsize\bfseries\textit{Chapter~1}}
\end{flushleft}
\vskip 0.45in
%\bfseries\scshape Chapter Title}}
%\bfseries\scshape Nuclear Forces at Short Distances}}
\bfseries\scshape Nucleon - Nucleon Interactions at Short Distances}}

%\author{\bfseries\itshape Authors\thanks{E-mail address: xxxx}\\
%Affiliations}
\author{\bfseries\itshape Misak M. Sargsian \thanks{E-mail address: sargsian@fiu.edu}\\
Florida International University, Miami, Florida, USA}

\date{}
\maketitle
\thispagestyle{empty}
\setcounter{page}{1}
% ------- [First Page Running Head] - place it immediately after title! ------
\thispagestyle{fancy}
\fancyhead{}
\fancyhead[L]{In: Book Title \\ 
Editor: Editor Name, pp. {\thepage-\pageref{lastpage-01}}} % needs \label{lastpage-01} on the last page.
\fancyhead[R]{ISBN 0000000000  \\
\copyright~2007 Nova Science Publishers, Inc.}
\fancyfoot{}
\renewcommand{\headrulewidth}{0pt}
%------------------------------------------------------------------------------

%\vspace{2in}

\begin{abstract}

Despite the  progress made in understanding the NN interactions at long distances based on effective field theories, 
the understanding of the dynamics of  short range  NN interactions remains as elusive as ever. 
One of the most fascinating properties of short range interaction is its repulsive nature which is responsible 
for the stability  of strongly interacting  matter. The relevant distances, $\le 0.5$~fm, 
in this case are such that one expects the onset of quark-gluon degrees of freedom with 
interaction being dominated by  QCD dynamics.   
We review the current status of the understanding of  the  QCD dynamics of  NN interactions at short distances, 
highlight  outstanding questions and outline the theoretical foundation of QCD description of hard NN processes.
We present  examples of how the study of the hard elastic NN interaction can reveal the symmetry structure of valence quark 
component of the nucleon wave function and how the onset of pQCD regime is correlated with the onset of color transparency 
phenomena in hard $pp$ scattering  in the nuclear medium.  The discussions show  how the  new experimental facilities can help 
to  advance the knowledge about the QCD nature of nuclear forces at short distances.
\end{abstract}

\noindent \textbf{PACS} 12.38.$-$t, 13.75.Cs, 21.30.$-$x.
\vspace{.08in} \noindent \textbf{Keywords:} Nuclear Forces, QCD, NN Interaction, Hard Processes.

% ------------ [Running Heads - for odd and even pages] - please insert it only on page 2!
\pagestyle{fancy}
\fancyhead{}
%\fancyhead[EC]{Authors}
\fancyhead[EC]{M.M.Sargsian}

\fancyhead[EL,OR]{\thepage}
%\fancyhead[OC]{Article Name}
\fancyhead[OC]{NN Interactions}

\fancyfoot{}
\renewcommand\headrulewidth{0.5pt} 
%------------------------------------------------------------------------------

\section{Introduction}
\label{sec1}
After the discovery of the Atomic Nuclei in 1911~\cite{Rutherford1} and the observation of  strong nuclear forces in 
1919~\cite{Rutherford2,Rutherford3},  the understanding  of  Nuclear Physics on a fundamental level  is still  an 
enduring intellectual challenge.   
We know why atomic nuclei are bound but we don't know why they are stable  in the way they are, especially heavy nuclei. 
The Stability Theorem~\cite{BW}  requires a repulsive interaction  in order to heavy nuclei not to collapse 
to the configurations with the radius of $\approx 0.5$~fm and  per nucleon binding energy of  $1500$~MeV!

It is remarkable that such a repulsive interaction was eventually found in the short range part of the   Nucleon-Nucleon~(NN) 
interaction, when  the analysis of the  NN  scattering data by Jastrow in 1951~~\cite{Jastrow} unambiguously demonstrated the 
existence of the repulsive interaction in the $NN$ systems at $\le 1$~fm.

The short-range nature of nuclear forces indicates that the bulk of the dynamics of atomic nuclei can be understood
from the dynamics of the NN  interaction.  Therefore it is not surprising that  special emphasis is given to 
the studies of NN interaction  throughout  the history of  nuclear physics.

If one studies the NN interaction starting from large distances (say $\ge 2$fm), one first observes 
a clear dominance of  the long range one-pion exchange  interaction. By decreasing the separation of  nucleons 
to  $1.2 - 1.5$~fm  one will observe the onset of the strong contribution due to two-pion exchange forces resulting 
in the tensor interaction. The  observed magnitude  of the tensor as well as  strong spin-orbit interactions, however, requires 
an addition of the vector component to  the exchanged forces   whose contribution  gradually increases with the decrease of 
NN separations and  dominates the overall interaction in the region of the repulsive core  at $\le 0.5$~fm.
The above picture allows us to identify three main regions (similar to Ref.~\cite{threereg})  for NN interactions which are  characterized  
by their  unique dynamics and therefore by their  specific  theoretical approaches in describing the NN forces.

{\bf Long to Intermediate Range:}
The biggest success in understanding the nuclear forces is achieved in describing long range and long to intermediate range pion 
exchange interactions based on  the effective theories (e.g. ~\cite{EFT1,EFT2,EFT3,EFT4}). The advantage of these theories is that 
they satisfy the symmetries of QCD Lagrangian and the interaction vertex is fixed by the mechanism of the Chiral symmetry breaking.
Effective theories can be extended to the intermediate-short range distances  by introducing contact interactions. Such an approach 
is  successful   in describing the intermediate - long range part of the NN interaction, however no dynamical insight can be  achieved on 
the structure of the short-range interaction.

{\bf Intermediate to Short Range:}
Intermediate to short range interactions are currently based fully on  phenomenological approaches. 
In One Boson Exchange Potential (OBEP) models  the vector meson exchanges  are  introduced with phenomenological 
coupling constants with the  assumption that they account for the field theoretical effects of 
multiple scattering, self interaction of exchanged mesons, etc.  (for review see Ref.~\cite{OBEP}).  
The other approach is in phenomenological  parameterization of the NN interaction  potential in the form of the sum of Yukawa type interactions (e.g. Ref.~\cite{NNpars}).
In both cases the parameters entering  the approximations  are  fixed by fitting  the calculations to the phase shifts of  interactions  
extracted from the analysis of the NN scattering data.  In fact the major success of these models are in part  due to the 
significant improvements in  quality and quantity  of  NN  phase-shift data  by SAID~\cite{SAID,SAIDupdt} and Nijmegen~\cite{Nijmegen} groups.

Despite the progress in quantitative description of NN interactions up to the lab kinetic energies of $350$~MeV 
the difficulty is however  in the fact the there is no clear cut constraints  that QCD can impose on  the dynamics of the interaction in 
these approaches.   

{\bf Short Range:}
Concerning  the short range ($< 0.5-0.7$~fm)  interaction, the most fascinating part of it is the repulsive core which we know is responsible 
for the stability of atomic nuclei.  This is the largely unknown part of the nuclear forces.   It is worth mentioning that the most modern 
phenomenological parameterizations~\cite{NNpars}  of  NN potentials still use the Wood-Saxon type function for the repulsive part of the
interaction  parameterized in  60's. What concerns  OBEP models, here the biggest inconsistency is that one considers the exchange of 
composite particles (bosons) which have  sizes comparable to or larger than the internucleon distances being considered. In the language of Feynman diagrams this situation is reflected in the fact that 
exchanged mesons are highly virtual (their virtuality being comparable with their masses).  In this respect it is worth to 
quote Richard Feynman who noticed that "In fact a "pion far off its mass shell" may be a meaningless - or at least highly complicated idea"~\cite{Feynman}.

{\bf Short Ranges and QCD:}
The interesting aspect of the short range repulsive interaction is that it operates at distances of $\le 0.5$~fm at which one expects full onset of 
the QCD degrees of freedom.  Therefore our expectation is  that the NN interaction  should  be  strongly related to the QCD dynamics at 
these distances. \\  
\indent The current experimental and theoretical situation in understanding the QCD dynamics of  NN interaction at short distances  is very 
incomplete. The  available experimental data indicate the rich dynamics at short-range  NN interaction. 
However, no new data were taken since 
mid 1990's after re-profiling the AGS at Brookhaven National Lab as an injector for 
Relativistic Heavy Ion Collider as well as completion of the  Saturne II project at Saclay~\cite{SaturneII}.   
The halt of the flow of experimental data    significantly restricted the possibility of further theoretical development.    
This situation is  expected to change in the near future  with the advent of  
high energy hadronic machines at J-PARC (Japan)~\cite{J-PARC}, PANDA (Germany)~\cite{PANDA} 
as well as the currently discussed NICA(Russia)~\cite{NICA} and HIAF(China)~\cite{HIAF} projects.

{\bf Outline:}
In this work we review briefly  the history of QCD studies of NN interactions (Sec.\ref{sec2}) and highlight the outstanding 
theoretical and experimental issues in understanding the QCD structure of high energy and momentum transfer 
(hard) NN interactions~(Sec.\ref{sec3}). 
We then present  the theoretical foundation of hard QCD processes relevant to NN interaction (Sections (\ref{sec5}-\ref{sec6}).
First we consider the amplitudes of NN interaction  in helicity 
basis, most appropriate  for high energy scattering. Then,  anticipating the  helicity conservation in  the perturbative regime of QCD 
in Sec.\ref{sec4} we 
discuss in detail the polarization observables in elastic NN interaction and derive several useful sum rules and equalities 
for polarization asymmetries  that can serve as a tool for identifying the onset of helicity conservation in the hard QCD regime.
In Sec.\ref{sec5} we discuss the  other signature of the  QCD regime of  hard NN Scattering - the  quark counting rule.   Sec.~\ref{sec6} 
elaborates the light-front dynamics of  hard NN scatterings, the definition of light-front wave function of nucleons as well as quark  helicity spinors in the light-front, giving 
us the main theoretical tools for studying hard NN interactions.  In Sec. \ref{sec7} we consider a few  examples of  the application
of the above mentioned theoretical  approaches in the quark-interchange model of NN interaction, comparing the results 
with the limited experimental data currently available.  Section \ref{sec8} summarizes the results and gives some outlook on other 
possibilities of studying NN systems at short distances.

\section{Brief History of QCD approaches to  Short Range NN Interaction}
\label{sec2}

The QCD approach in studies of  high momentum and energy transfer NN  interactions  revealed a very rich picture of  strong interactions, 
 introducing  completely new subjects into  the  discussion of nuclear forces such as dimensional counting rules, helicity conservations, minimal Fock components, hidden color,  etc...

{\bf Quark-Gluon Interactions and NN Scattering:}
First attempts to understand short range properties of $NN$ interaction within QCD were made in the 1970's~\cite{BBL73,BBL75, Landshoff}, 
in which the phenomenology of $NN$ interaction was confronted with the quark-interchange and three gluon exchange mechanisms of 
hadronic interactions.   The real boost in studies of hard hadronic processes in general and $NN$ interaction in particular came 
after the discovery of dimensional counting rules~\cite{MMT,BF1,BF2}  for fixed  large center of mass angle  
high energy processes.  The counting rules gave new observables  to explore in hard  processes  
such as  the exponents of the energy dependences of scattering cross sections.
The most intriguing part of it was  that such dependences were naturally predicted within QCD in the picture of hard gluon exchanges 
between current quarks in hadrons.  

The  another feature of  the  processes proceeding through the hard gluon exchanges  was the prediction of helicity conservation and new 
relations between different spin observables  in the hard hadronic processes~\cite{FG,BLC}.   Naturally, the most accessible hadronic processes 
were the $NN$ scatterings and  there were several dedicated experiments 
on measuring different spin observables in hard NN processes. However, the experimental results were at best controversial~\cite{xpol1,xpol2} 
with the  most "damaging" being the measurement of the $A_{nn}$  polarization~\cite{Crabb,Crosbie}, which observed that for $90^0$ center 
of mass elastic $pp$ scattering at $p_{Lab} = 11.75$~GeV/c,  the cross section of  protons scattering with parallel spins normal to the scattering 
plane is four times the one with the antiparallel spins.

{\bf The Quark-Gluon Wave Function of the Nucleon:}
To calculate the NN scattering cross section within QCD, one needs, as an input, the quark-wave function of the nucleon.  However, this 
wave function is poorly known due to the immense complexity of the QCD picture of  the nucleon  in which the number of  
quark and gluon constituents are not conserved (for the recent review on bounds rates in QCD see Ref.\cite{Hoyer}). 

On the other hand, the advantage of high energy and momentum transfer reactions is 
that they 
allow a successful application of Fock-component approach to describing the wave function of a nucleon.   
This approach allows us to start with the  minimal Fock component of the nucleon consisting of only three valence quarks and 
treat the higher order Fock-components as  corrections since they are suppressed by additional inverse orders of 
the invariant  momentum transfer.

Concentrating  the minimal-Fock component of the nucleon, the next question is what symmetry of valence quarks one 
needs to consider in the wave function of the nucleon.  Here the most natural is to consider the SU(6) symmetry based on the fact that 
the very same symmetry seems to be  realized in constituent quark model which describes the baryonic spectrum reasonably well.
Indeed the SU(6) symmetry  was  one of the the first approaches in describing hard NN processes~\cite{FG,BLC}.  
Also the constituent quark formalism 
described reasonably well several properties of intermediate energy NN scattering within  the SU(6) symmetry~\cite{Isgur}.  
\\

{\bf QCD and Short-Range Structure of the NN-Bound System:}
The  third aspect  which is worth to emphasize  in relation to  studies of NN interaction at short distances is 
the dynamics of the NN-bound system probed at short distances.   The deuteron studies opened up a new 
realm in studies of QCD dynamics of strong forces.  It was realized in Refs.~\cite{Harvey,Obukhovsky:1982ci,BLL1,BLL2,Kusainov:1991vn}, 
that  the fact, that  the deuteron is a colorless 6-valence-quark  system, creates an additional  possibility for existence of color-octet 
three-quark (3q)  states that combine into color-singlet 6-quark combinations (referred as hidden color states).  
The calculations indicate that there is a  substantial hidden color component  in 
the NN bound system at short distances in which the 6q system becomes a  relevant degree of freedom.

The notion of the hidden-color component gave a new possible meaning  to the NN repulsion which  can be in part  due to  orthogonality between 
the initial two color octet  and final  two color singlet nucleons.    
The other implication of the hidden color component is the 
prediction of the large  contribution from  the $\Delta-\Delta$ component in the NN interaction following  from  
the decomposition of the color-singlet 6-quark system.

\section{Outstanding Questions  of  NN Interactions at Short Distances.}
\label{sec3}

There are many unresolved questions which are related to the nucleon-nucleon interactions at short distances 
with implications ranging from particle physics to astrophysics.  
With the expectation of a  new generation of  experimental 
studies of hard NN processes in  the near future,  it is worth enumerating several of the 
problems which can be  addressed in  the experiments at facilities mentioned earlier (Sec.\ref{sec1}).

{\bf Persistence of the Nucleonic Degrees of Freedom:}
Recent observations of large ($\sim 2M_{Sun}$)  neutron star masses~\cite{NSmass}  indicates existence  of  rather unreasonably 
stiff equation of  state of  the nuclear matter, which is related to the  persistence of the nucleonic degrees of freedom~\cite{PH}
 at such high densities in which one expects plenty of inelastic transitions and strong overlap of nucleon wave functions.  Such persistence 
 is also  observed in  probing short-range proton-neutron correlations in the nuclei~\cite{isosrc,eip4} in which the theoretical analysis~\cite{srcrev} 
 shows that for up to $\le 1$~fm separations, the  NN system  has no apparent non-nucleonic component, consisting almost entirely from 
 proton and neutron.

It is interesting that this observation also has its reflection  in the  modification of partonic distributions of bound nucleons  
in the nuclear medium (EMC effect). Here, the recent analysis~\cite{FSEMCnew}  indicates a rather  small modification  of nucleons 
in the  medium of heavy nuclei, which seems puzzling.  

Such a persistence of the nucleons in the high density nuclear environment  can be due to 
the short range repulsion, since the attractive interaction  will make the composite system very responsive to medium modifications.

 \begin{figure}[ht]
%\vspace{-0.4cm}
\centering\includegraphics[scale=0.5]{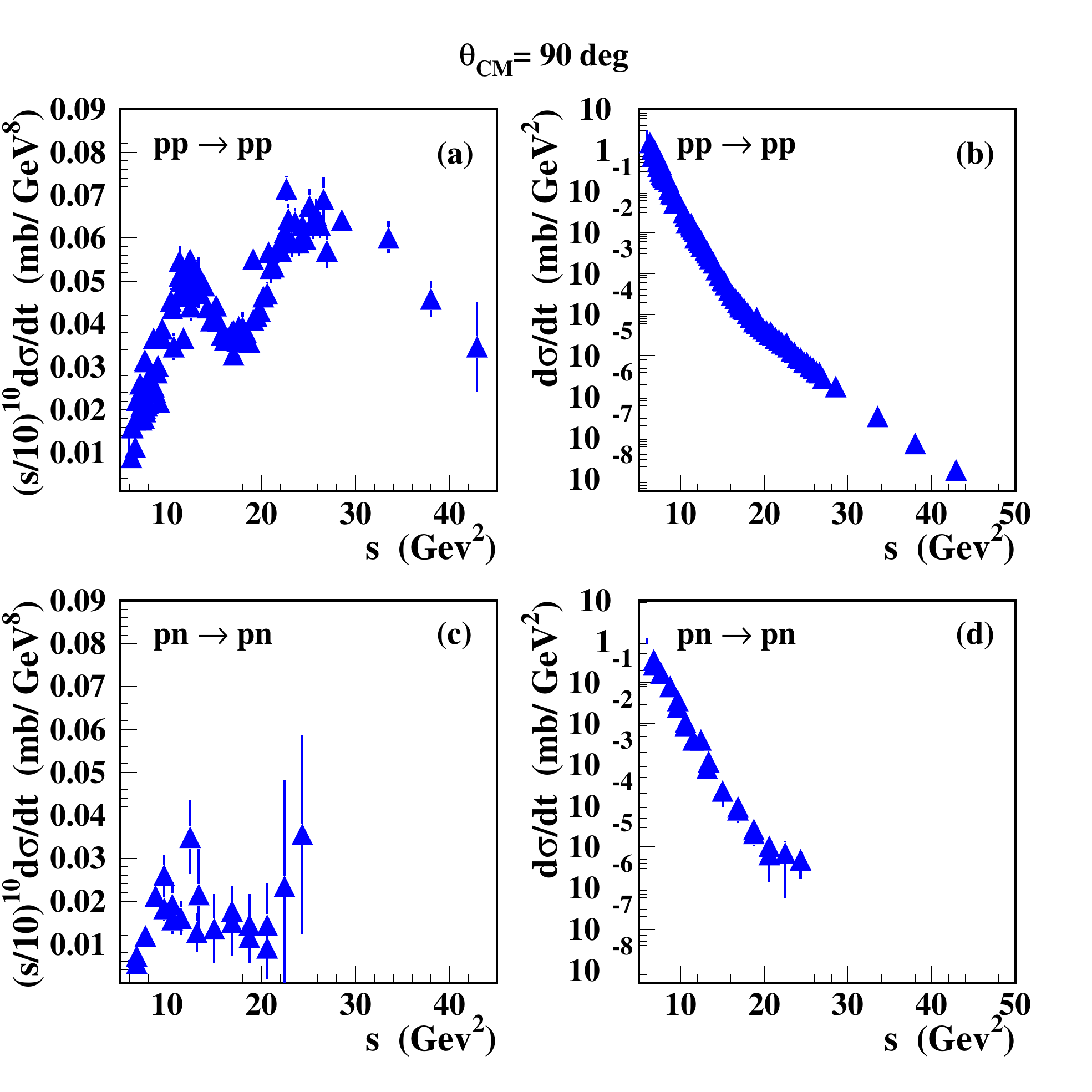}
%\vspace{-0.4cm}
\caption{The invariant energy dependence of elastic $pp$ and $pn$ differential cross sections  unweighted (b) (d) and weighted 
by $s^{10}$ factor (a), ({c}). The experimental data are from Ref.~\cite{hepdata,Allaby,Akerlof,pnexp1,pnexp2}.}
\label{s10}
\end{figure}

{\bf Oscillatory Energy Dependence of the Scaled Cross Section of Hard NN Interaction:}
Even though the fixed angle high energy $NN$ scattering demonstrated rather unambiguous $s^{-10}$ invariant energy dependence,  in 
agreement with quark counting rules~\cite{BF1,MMT,BF2}, the more careful analysis revealed a violation of this dependence in a rather peculiar form:
while  the quark-counting rule predicts energy independence of $s^{10}{d\sigma\over dt}$, the data clearly demonstrate an oscillatory
energy dependence of this combination (Fig.1). These regularities   observed in this residual energy dependence are striking 
and  despite several intriguing explanations~\cite{RP,BdTer}, their origin  is not understood. 
In Ref.~\cite{RP} it is assumed that in addition to the quark-interchange 
mechanism of  hard NN scattering which is sensitive to the quark-gluon interaction at very short distances, the  three-gluon exchange mechanism
contributes to the overall scattering amplitude.  The latter (often referred to as Landshoff mechanism~\cite{Landshoff})  does 
not require a quark-interchange, thus, the short distances are not necessarily involved in the NN interaction. Because of 
the possible Chromo-Coulomb  phase shift, the two amplitudes enter with the relative phase that can be tuned to describe the  observed 
energy oscillations.

In Ref.~\cite{BdTer} it is assumed that the oscillations are due to the interference of the 
quark-interchange pQCD 
amplitudes and the  amplitudes of  near-threshold  $s$-channel strangeness and charm  productions 
with possible resonating configurations having quantum numbers of $J=1$, $L=1$ and $S=1$.  
With the relative phases of the  pQCD and resonating amplitudes  
taken as a free parameter, it was possible to describe the observed energy dependences reasonably well.

Overall, in both approximations the assumption is that the pQCD short-range quark-interchange amplitude  is interfering 
with another  amplitude which represents a  spatially extended system.

Concluding this discussion,   it is worth noting that the  irregularity in the energy dependence  is on a level of $60\%$   in the region 
where the magnitude of the  hard elastic $pp$ cross section changes by  {\em eight} orders of magnitude (Fig.\ref{s10}(b)). 
As Figs.\ref{s10}(b) and ({c}) show, the same is apparently true for $pn$ hard elastic scattering.

\begin{figure}[ht]
%\vspace{-0.4cm}
\centering\includegraphics[scale=0.45]{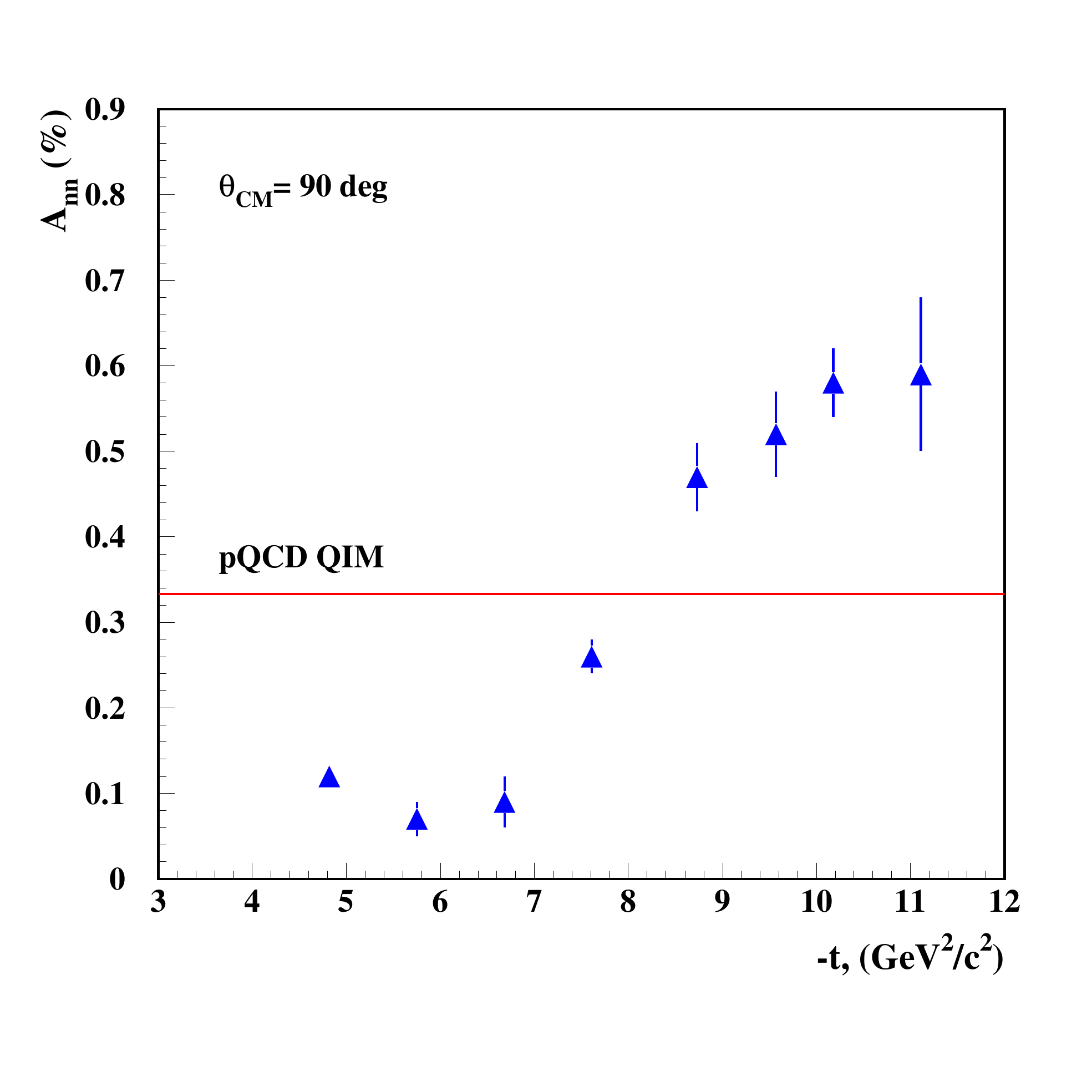}
\vspace{-0.4cm}
\caption{The $-t$ dependence of the analyzing power parameter  ($A_{nn}$)  of  hard elastic pp scattering. Data are from Ref.~\cite{Crosbie}.
The curve labelled as "pQCD QIM" is the prediction within pQCD quark interchange model (Sec.7).}
\label{ann}
\end{figure}

{\bf Anomalous Polarization Asymmetries  in Hard NN Scattering:}
One of the  biggest challenges in understanding the QCD dynamics of hard $NN$  interactions came from the observation of large  out-of-plane 
polarization effects at $90^0$ cm scattering  for incoming proton momenta as large as $p_{lab} = 11.75$~GeV/c (Fig.\ref{ann}).   The surprise in these results is that 
the experiment found that the elastic cross section of the protons that scatter with their spins  parallel and normal to the scattering plane 
is  almost  {\em four} times larger than the cross section  when the spins are antiparallel.   

The $A_{nn}$ experiment resulted in a  flurry of theoretical activity (see e.g.~\cite{FG,BLC,ACM,Lipkin,Bourrely,Preparata,Troshin,AKP,GR}.
 However  the origin of large asymmetry and its energy dependence are still unresolved.

The implication of this result is  that, in order to generate such 
a large polarization effects, one needs to have a large contribution from the double-spin flip amplitude $\phi_2$  or negligible contribution from  
the helicity conserving $\phi_1$ amplitude (see Sec.\ref{sec7}).  However, the challenge here is that in the hard QCD regime,  the  $\phi_2$ 
is the most suppressed and the $\phi_1$ is the largest.   Another challenge that the experiment posed was (again)   the "oscillatory"  energy dependence of the 
asymmetry (Fig.\ref{ann}). Its difference from the  basically energy independence of  the pQCD quark-interchange prediction  
is rather striking (Fig.\ref{ann}),

The above situation indicates that, whatever the  origin  of this large asymmetry  is, it is  related to the contribution of 
a spatially extended configuration into the  process of the $pp$  scattering. 
The oscillatory energy dependence indicates that there is an interference between short and long range phenomena.   For example, the 
earlier mentioned charm threshold enhancement effect with the intermediate $J=1$, $L=1$, $S=1$ quantum state results in $\phi_1=0$ and its 
interference with the background pQCD QIM  process can in principle generate the energy dependence of Fig.\ref{ann} (see Ref.~\cite{BdTer}).

{\bf Color Transparency in Hard  $NN$ Scattering:} One of the most remarkable predictions of QCD is the existence of color transparency 
phenomena in hard  processes taking place in  the nuclear environment. The possibility of producing small-sized color-neutral object 
in hard processes  will be evidenced by the reduced absorption  of hadrons  in propagating through the nuclear medium.  The interesting QCD feature 
is  that such a neutral object can be produced both for mesons and baryons.   In large momentum transfer, the small distances involved 
in quark-gluon interaction  will require that the quarks in the minimal Fock component of  hadron wave function will populate 
distances $\le 1/\sqrt{-t}$. Such configurations will have small color dipole moments  which will result in reduced strong interaction in the 
nuclear medium~\cite{CTB,CTMul}. These expectation resulted  in  tremendous theoretical and experimental activities in mid 1990's 
(for review of the subject see e.g. Refs.~\cite{CTFMS,CTRP}).  
The realization that these small-sized configurations are not  eigenstates of QCD Hamiltonian of
free hadrons and  once produced at finite energies  they will evolve to  normal size hadrons,  suggested  that the experimental verification of color transparency phenomena  is more complex than first expected(see e.g. Ref.~\cite{FLFS,Kopeliovich}). 

\begin{figure}[ht]
%\vspace{-0.4cm}
\centering\includegraphics[scale=0.45]{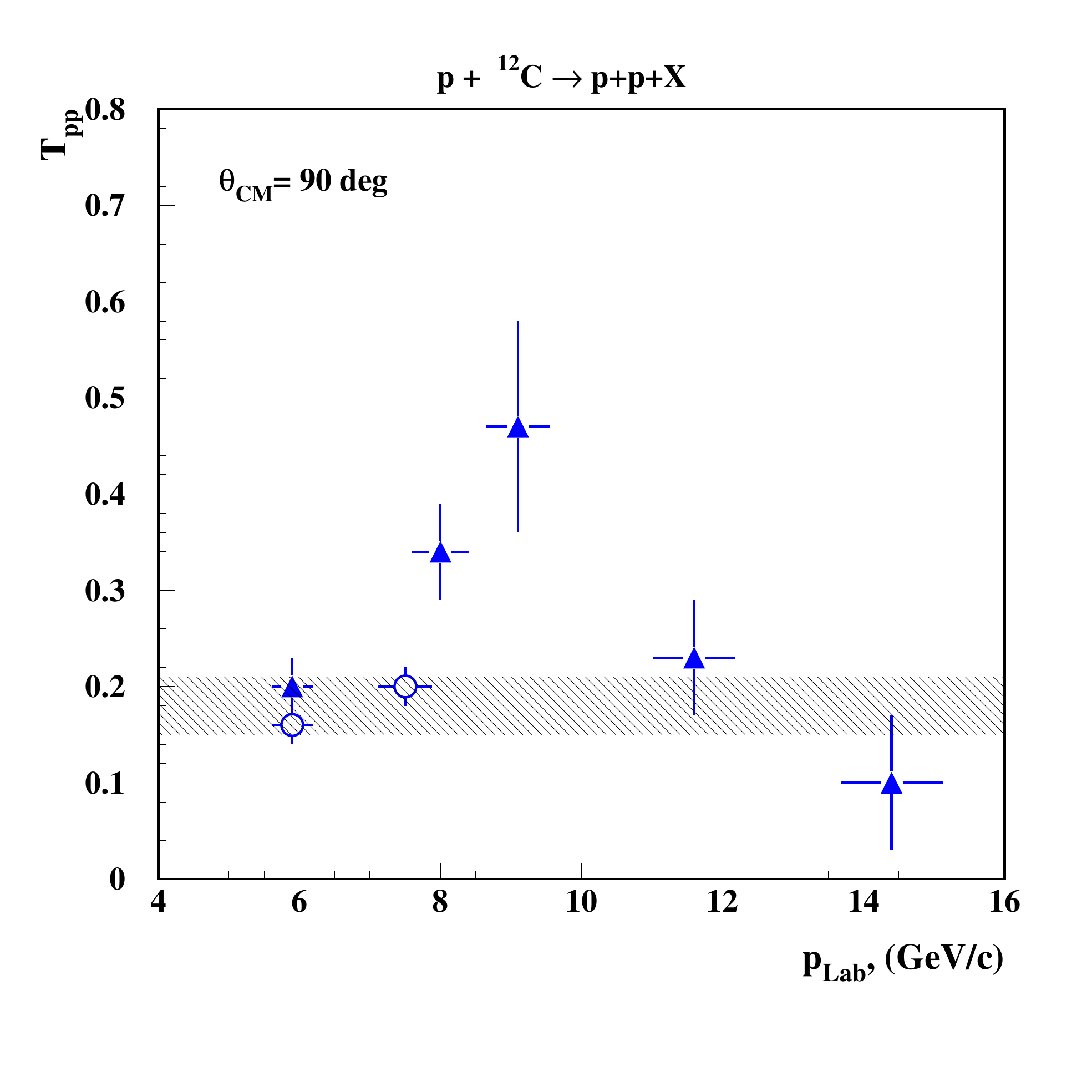}
\vspace{-0.4cm}
\caption{Dependence of the nuclear transparency  parameter on the momentum of the incoming proton. 
Data are from Ref.~\cite{Mardor,Aclander}. Shaded area represents the calculation based on the Glauber theory~\cite{FSZ,FMSS} in which 
the transparency is defined  based on the total cross section of $NN$ scattering.}
\label{ct}
\end{figure}

While the Color Transparency phenomenon or the reduction of  the absorption in the nuclear medium is  observed  
for the hard production of $q\bar q$ systems~\cite{ct2jets,ctrho,ctpi}, the similar effect  is still elusive for a $qqq$ system.  
The first experiments based on the hard $pp\to pp$ elastic scattering from nuclei~\cite{Carroll,Mardor,Aclander} demonstrated 
another  "oscillation" feature - this time in the proton beam momentum dependence of  the nuclear absorption parameter (Fig.\ref{ct}).

The interesting feature of these data is that initially the transparency grows with the proton momentum  above the values 
expected from Glauber theory, and then after $p_{lab} = 9$~GeV, it falls back leading to transparency 
values apparently below the  Glauber prediction at $p_{Lab} \approx 14$~GeV.  The latter indicates that whatever propagates 
at these energies in the nuclear medium has total cross section larger than that of  the  $NN$ scattering.
  
 {\bf Oscillations Superimposed:} If now one superimposes Figs.(\ref{s10}),(\ref{ann}) and (\ref{ct}) recalculating them for 
 invariant energy dependences at given $\theta_{cm}=90^0$ one observes rather interesting correlations between these three different 
 measurements.
 
 \begin{figure}[ht]
%\vspace{-0.4cm}
\centering\includegraphics[scale=0.45]{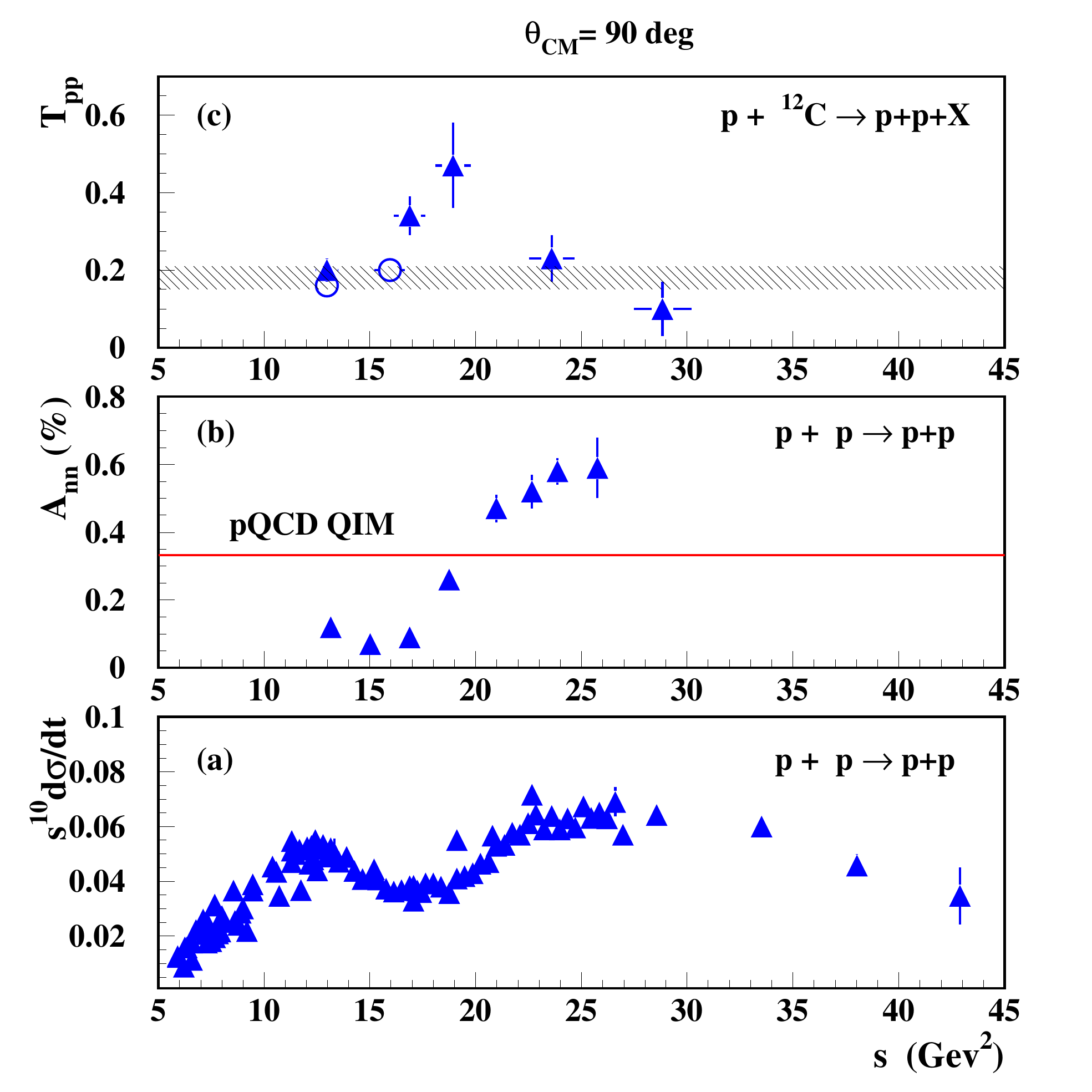}
\vspace{-0.4cm}
\caption{Dependence of the nuclear transparency ({c}),  asymmetry, $A_{nn}$ (b) and $s^{10}$ scaled elastic $pp$ cross section (a)  on the invariant energy $s$.}
\label{ct_ann_pp}
\end{figure}
First, the data suggest an inverse correlation between  the $s^{10}$ weighted $pp$ cross section (Fig.\ref{ct_ann_pp}(a))  and nuclear transparency 
$T_{pp} ($Fig.\ref{ct_ann_pp}({c})).
This correlation strongly indicates  that  the broad peak in the $pp$ cross section at $s\approx 25$~Gev$^2$ is related to the spatially 
large configuration which is strongly absorbed in the nuclear medium as indicated by  the drop of the $T_{pp}$ at the same $s$.  
Thus, the observed correlation suggests  that  the  nuclei filter the large-size  component from the amplitude of hard $pp$ scattering~\cite{CTRP}. 
This picture is reinforced when one compares  the $s$-dependences  of  the  same 
$s^{10}$ weighted $pp$ cross section (Fig.\ref{ct_ann_pp}(a))  and  the polarization asymmetry, $A_{nn}$ (Fig.\ref{ct_ann_pp}({b})).  
As Fig.\ref{ct_ann_pp}({b})) shows,
at $s\approx 25$~Gev$^2$ the $A_{nn}$ produces the largest asymmetry  well above the pQCD prediction~\cite{BdTer} (see Sec.\ref{sec7}), while
the  pQCD prediction is closest to the $A_{nn}$ data at $s\sim 18-20$~GeV$^2$. The interesting result here is that, 
at the latter energies  ($s\sim 18-20$~GeV$^2$)   the onset of Color Transparency (less attenuation) is observed in 
Fig.\ref{ct_ann_pp}({c})). Such a  correlation  between the agreement of the pQCD prediction of $A_{nn}$ with the data and the 
increase of  the nuclear transparency, $T_{pp}$,  seems  to validate the  main premise of  the  Color Transparency 
prediction, that the small-sized configurations are produced in  the pQCD regime which propagate in the nuclear medium with 
less absorption than ordinary hadrons.

{\bf Diquark Signatures of Nucleons:} There is a rather rich phenomenology about the quark-diquark   structure of the nucleon. This 
phenomenology follows from the hadron spectroscopy, deep-inelastic scattering, baryon form-factors as well as exclusive
 hadron interactions (e.g. ~\cite{Anselmino,Wilczek}).  The diquark  structure reveals itself both as a specific symmetry of quark wave function of the 
 nucleon as well as dynamical component that contributes to the scattering process (for example through the diquark
 exchange).  However, the impossibility of formulation of diquark-structures from the first principles of QCD and its inclusion into the 
 self-consistent  evolution equation scheme makes the issue of diquarks an open question in  NN scattering  physics.

{\bf Hidden Color Component in the NN-System:} The hidden color component is one of the unique QCD effects in NN interaction 
that can not be imitated by any baryons degrees of freedom.  The hypothesis of the color-neutrality of  the observed strongly interacting 
composite systems  introduces the possibilities for a new reality in which  baryonic systems with high degree of compositeness 
(one example is NN system) contain  explicitly colored  three-quark clusters that combine into a  color  neutral object. 
One of the best examples is the contribution of two colored baryons, $N_c$,  into the colorless NN system~\cite{Harvey,BLL1,BLL2}.
While existence of such hidden-color components are  accepted within QCD  there is no  clear experimental evidence yet for  
such components.

\section{Helicity Amplitudes  and Observables of  Elastic NN Scattering}
\label{sec4}

In the following three sections an outline of the theoretical approach is given, which is most suitable for the description of the 
hard $NN$ scattering in QCD.

The  natural framework in describing $NN$ scattering amplitudes  at high energies is the helicity description of the quantum states of 
the nucleon.  This representation has the obvious advantage of satisfying the boost invariance of the   observable quantities as well 
as allowing us to define their asymptotic limits in the case of helicity conservation, which is the case in  the pQCD regime of the scattering.

For exclusive, $1+2\rightarrow 3+4$ scattering of spin-half particles, in general, one has a total of 16 helicity amplitudes 
$\langle \lambda_3\lambda_4\mid M \mid \lambda_1\lambda_2\rangle$, where
$\lambda_i$ describes the helicity state of the $i$'th nucleon having $+{1\over 2}$ and $-{1\over 2}$ eigenvalues for which we also 
use the notations $+$ and $-$ respectively\footnote{Subscripts "1", "2", "3" and "4" hereafter represent the incoming, target, scattered and 
recoil nucleons respectively.}.
Parity conservation and  time reversal invariance of strong interaction as well as the application of 
generalized Pauli principle (or Isospin symmetry) introduces the following relations  between helicity amplitudes 
of the elastic NN scattering (see e.g. ~\cite{Perl,Bystricky}):
\begin{eqnarray}
\langle -\lambda_3-\lambda_4\mid M \mid -\lambda_1-\lambda_2\rangle   & = &  (-1)^{(\lambda_1-\lambda_2 - (\lambda_3-\lambda_4))} \langle \lambda_3\lambda_4\mid M \mid \lambda_1\lambda_2\rangle \nonumber \\
\langle \lambda_1 \lambda_2\mid M \mid \lambda_3 \lambda_4\rangle      & = &  (-1)^{(\lambda_1-\lambda_2 - (\lambda_3-\lambda_4))} \langle \lambda_3\lambda_4\mid M \mid \lambda_1\lambda_2\rangle\nonumber \\
\langle \lambda_4 \lambda_3\mid M \mid \lambda_2 \lambda_1\rangle      & =  & (-1)^{(\lambda_1-\lambda_2 - (\lambda_3-\lambda_4))} \langle \lambda_3\lambda_4\mid M \mid \lambda_1\lambda_2\rangle.
\end{eqnarray}
These relations reduce  the number of independent elastic $NN$  helicity amplitudes  to five  which we define as follows:
\begin{eqnarray}
&& \phi_1  \equiv      \langle ++\mid M \mid ++\rangle =  \langle --\mid M \mid --\rangle \nonumber  \\
&& \phi_2   \equiv     \langle ++\mid M \mid --\rangle =  \langle --\mid M \mid ++\rangle  \nonumber  \\
&& \phi_3   \equiv    \langle +-\mid M \mid +-\rangle =  \langle -+\mid M \mid -+\rangle  \nonumber  \\
&& \phi_4    \equiv     \langle +-\mid M \mid -+\rangle =  \langle -+\mid M \mid +-\rangle  \nonumber \\
&& \phi_5   \equiv    \langle ++\mid M \mid +-\rangle =  \langle -+\mid M \mid --\rangle =   \langle --\mid M \mid +-\rangle  = \langle -+\mid M \mid ++\rangle   = 
  \nonumber  \\
&& \ \    -\langle --\mid M \mid -+\rangle = - \langle +-\mid M \mid ++\rangle =   -\langle ++\mid M \mid -+\rangle  = -\langle +-\mid M \mid --\rangle, 
\nonumber \\
\label{phis}
\end{eqnarray}
where $\phi_i$'s are functions of invariants, $s$ and $t$, or can be alternately presented as a function of $s$ and the scattering angle, $\theta_{cm}$ in the 
center of mass  reference frame. From the above relations it follows that for isotriplet states, particularly for  elastic $pp$ scattering at $\theta_{cm}=90^0$, 
one has $\phi_5^{pp}(\theta_{cm} = 90^0) = 0$.

\subsection{Practical Observables  for  Hard Elastic  $NN$ Scattering}

There is one unpolarized observable which is the cross section of elastic $NN\rightarrow NN$ scattering defined as:
\begin{equation}
{d\sigma\over dt} = {1\over 16\pi} {1\over s(s-4m_N^2)} \sigma_0 \ \ \ \mbox{ with} \ \ \ 
\sigma_0 = {1\over 2} (|\phi_1|^2 +  |\phi_2|^2 + |\phi_3|^2 + |\phi_4|^2 + 4|\phi_5|^2)
\end{equation}
and a total of  256 possible polarization observables in the center of mass reference frame.
However, the application of the above mentioned parity conservation, time reversal invariance and generalized 
Pauli principle reduces this number to 24 independent  polarization observables (see e.g.~\cite{Bystricky}).  

Considering the fact that  the most of the  possible $NN$ scattering experiments in the  near future will be performed at fixed-target kinematics, we 
will consider  the polarization observables for the  lab frame kinematics.  For this purpose we first 
define the momenta of incoming, target, scattered and recoil nucleons  as ${\bf p_1}$, ${\bf p_2} (= 0)$, ${\bf p_3}$ and ${\bf p_4}$ respectively.
Then in addition to three directions defined by above vectors:
\begin{equation}
{\bf \hat p_1} = {{\bf p_1} \over |p_1|},   \ \ \ {\bf \hat p_3} = {{\bf p_3} \over |p_3|},   \ \ \  { \bf\hat p_4} = {{\bf p_4} \over |p_4|}
\end{equation} 
one has  four following vectors:  The one,  $n$,  is transverse to the scattering plane:
\begin{equation}
{\bf n } = {{\bf p_1}\times {\bf p_3}\over |{\bf p_1}\times {\bf p_3}|}
\end{equation}
and three remaining define in-scattering-plane directions:
\begin{equation}
{\bf s_1 } = {{\bf n}\times {\bf p_1}\over |{\bf n}\times {\bf p_1}|}, \ \ \ {\bf s_3 } = {{\bf n}\times {\bf p_3}\over |{\bf n}\times {\bf p_3}|}, \ \ \ 
 {\bf s_4 } = {{\bf n}\times {\bf p_4}\over |{\bf n}\times {\bf p_4}|}.
\end{equation}
Based on these directions we define a  polarization observable
\begin{equation}
X_{3,4,1,2},
\end{equation}
where subscripts $1,2,3,4$ correspond  to the directions of the polarization of the initial, target, scattered and recoil nucleons. Each 
of these subscripts can have  polarizations along all the above defined directions.

\medskip
\medskip
Furthermore, we will present only those polarization asymmetry observables which are experimentally most feasible (or practical) categorizing them  
by their {\em degree} of polarization.  Also at the right hand side of the each polarization asymmetry observable we give the corresponding 
expression  in the limit of  helicity conservation. \\

\noindent{\bf Single Polarization:}\\ There is only one nonzero  single polarization observable defined by the direction of $\bf n$:
\begin{equation}
X_{n,0,0,0}  = X_{0,n,0,0} = X_{0,0,n,0} = X_{0,0,0,n} = - {1\over \sigma_0} Im\left[\phi_5^\dagger (\phi_1 + \phi_2 + \phi_3 - \phi_4)\right] 
\rightarrow 0.
\label{Py}
\end{equation}

\noindent {\bf Double Polarizations:}\\
(a) Out-of-plane initial polarization asymmetry  and polarization transfer observables:
\begin{eqnarray}
X_{0,0,n,n}  & =  & X_{n,n,0,0} = {1\over \sigma_0}\left( Re\left[\phi_1^\dagger \phi_2 - \phi_3^\dagger \phi_4\right] + 2|\phi_5|^2 \right)
\ \ \rightarrow - {1\over \sigma_0}Re\left[\phi_3^\dagger \phi_4\right] \nonumber \\
X_{n,0,n,0}  & =  &  X_{0,n,0,n}    = {1\over \sigma_0} \left(Re\left[\phi_1^\dagger \phi_3 - \phi_2^\dagger \phi_4\right] + 2|\phi_5|^2 \right)
\ \ \rightarrow  \ \ \ {1\over \sigma_0} Re\left[\phi_1^\dagger \phi_3\right] \nonumber \\
 X_{0,n,n,0} & =  &  X_{n,0,0,n}  = -{1\over \sigma_0}\left( Re\left[\phi_1^\dagger \phi_4 - \phi_2^\dagger \phi_3\right] + 2|\phi_5|^2 \right)
 \rightarrow -{1\over \sigma_0}Re\left[\phi_1^\dagger \phi_4\right] 
 \label{IniY}
\end{eqnarray}
 
\noindent (b) Scattering plane initial state polarization asymmetry observables:
\begin{eqnarray}
 &  & X_{0,0,s_1,s_1}   =       {1\over \sigma_0} Re\left[\phi_1^\dagger \phi_2 +  \phi_3^\dagger \phi_4\right]
 \rightarrow {1\over \sigma_0} Re\left[\phi_3^\dagger \phi_4\right]   \nonumber \\
&  & X_{0,0,s_1,\hat {p}_1}   =    X_{0,0,\hat {p}_1,s_1} = {1\over \sigma_0} Re\left[\phi_5^\dagger( \phi_1  + \phi_2 - \phi_3 + \phi_4)  \right]  
\rightarrow 0 \nonumber \\
&  & X_{0,0,\hat{p}_1,\hat {p}_1}   =   - {1\over 2\sigma_0} \left(|\phi_1|^2 +  |\phi_2|^2 - |\phi_3|^2 -  |\phi_4|^2  \right) 
\rightarrow  - {1\over 2\sigma_0} \left(|\phi_1|^2 - |\phi_3|^2 -  |\phi_4|^2  \right) \nonumber \\
\label{IniS}
\end{eqnarray}
 
 \noindent ({c}) Helicity transfer and double helicity final state observables:
\begin{eqnarray}
&  & X_{\hat{p}_3,0,\hat{p}_1,0}  = {1\over \sigma_0}\left(-Re\left[\phi_5^\dagger(\phi_1-\phi_2+\phi_3+\phi_4)\right]\sin{\theta_3}  \right.   \nonumber \\
   &  & +  \left. {1\over 2}\left[|\phi_1|^2 - |\phi_2|^2 + |\phi_3|^2- |\phi_4|^2\right]\cos{\theta_3}\right) 
 \rightarrow {1\over 2 \sigma_0}\left[|\phi_1|^2 + |\phi_3|^2- |\phi_4|^2\right]\cos{\theta_3}  \nonumber \\
&  & X_{0,\hat{p}_4,\hat{p}_1,0}  = {1\over \sigma_0}\left(-Re\left[\phi_5^\dagger(-\phi_1+\phi_2+\phi_3+\phi_4)\right]\sin{\theta_4} \right.  \nonumber  \\
 &  & -  \left. {1\over 2}\left[-|\phi_1|^2 + |\phi_2|^2 + |\phi_3|^2- |\phi_4|^2\right]\cos{\theta_4}\right) 
 \rightarrow {1\over 2\sigma }\left[\phi_1|^2 - |\phi_3|^2+ |\phi_4|^2\right]\cos{\theta_4}
 \nonumber \\
&  & X_{\hat{p}_3,\hat{p}_4,0,0}  = {1\over \sigma_0}\left({1\over 2}\left[|\phi_1|^2 + |\phi_2|^2 - |\phi_3|^2- |\phi_4|^2\right]\cos{\theta_3}\cos{\theta_4}
\right.
 \nonumber \\
&  &  \ \   + Re\left[\phi_1^\dagger \phi_2 + \phi_3^\dagger \phi_4\right]\sin{\theta_3}\sin{\theta_4} 
% \nonumber  \\
%& & \ \ 
\left.  -Re\left[\phi_5^\dagger(\phi_1+\phi_2-\phi_3+\phi_4)\right]\sin(\theta_3-\theta_4)   \right) \nonumber \\
&  & \ \ \rightarrow {1\over \sigma_0} \left({1\over 2}\left[|\phi_1|^2 - |\phi_3|^2- |\phi_4|^2\right]\cos{\theta_3}\cos{\theta_4}
 + Re\left[\phi_3^\dagger \phi_4\right]\sin{\theta_3}\sin{\theta_4}\right)  ,\nonumber \\
\label{IniT}
\end{eqnarray}
where $\theta_3$ and $\theta_4$ are the lab scattering angles of scattered "3" and recoil "4" nucleons.\\

\noindent {\bf Triple Polarizations:}
Here we consider the situations in which the transverse polarization of   only one of the final nucleons are measured in 
addition to the polarizations of initial nucleons:
\begin{eqnarray}
 X_{n,0,s_1,s_1}  & =  & X_{0,n,s_1,s_1} = -X_{n,0,\hat {p}_1,\hat {p}_1}  = -X_{0,n,\hat{p}_1,{\hat p}_1}  = 
 \nonumber \\
 & &      = 
 -  {1\over \sigma_0} Im\left[\phi_5^\dagger (\phi_1 + \phi_2 - \phi_3 + \phi_4)\right]    \ \ \rightarrow 0 \nonumber \\
X_{n,0,\hat {p}_1,s_1}  & = & X_{0,n,\hat {p}_1,s_1}  =  \ \ \ {1\over \sigma_0} Im\left[\phi_1^\dagger \phi_4 - \phi_2^\dagger \phi_3\right]  
\ \ \rightarrow 0 \nonumber \\
X_{n,0,s_1,\hat {p}_1}  & = & X_{0,n,s_1,\hat {p}_1}  =  -{1\over \sigma_0} Im\left[\phi_1^\dagger \phi_3 - \phi_2^\dagger \phi_4\right]  \ \ \rightarrow 0
\label{Triplepol}
\end{eqnarray}

\subsection{Sum Rules and  Equalities  for Scattering of Identical Particles} 

The polarization asymmetry observables, defined above, allow us to extract all the individual helicity amplitudes of $NN$ elastic scattering.
In practice it is less likely that 
all of these observables will be experimentally available at some point. However, several sum rules and equalities  between different observables 
(which have better chance of  being experimentally  measured)  will allow us to ascertain the extent of the polarization  in the hard scattering 
regime.

First, it is worth emphasizing  that, without invoking the assumption of  helicity conservation,  for identical particles 
one can demonstrate the  existence of  large asymmetries  for initial state polarizations at any  given high energy scattering 
at $\theta_{cm} = 90^0$.  For this, one observes that for identical particles at $\theta_{cm} = 90^0$ elastic scattering, (i.e. $pp\rightarrow pp$ scattering)  
due to the generalized Pauli principle, the amplitudes  with a single helicity flip are zero ($\phi_5 = 0$).  From the same principle  it follows that 
$\phi_4(\theta_{cm} = 90^0) = - \phi_3(\theta_{cm} = 90^0)$.
These properties of helicity amplitudes   allow us to derive the following relations between the above  defined polarization observables 
at $\theta_{cm} = 90^0$  elastic scattering of  protons:  The first relation is for double polarized initial states~\cite{FG,BLC}:
\begin{equation}
X^{pp}_{0,0,n,n} - X^{pp}_{0,0,s_1,s_1} - X^{pp}_{0,0,\hat{p}_1,\hat{p}_1}  = 1,
\label{psum}
\end{equation}
where for these polarization observables in the literature often the following notations are used:
$A_{nn} \equiv X_{0,0,n,n}$, $A_{ss}  \equiv X_{0,0,s_1,s_1} $ and $A_{ll}\equiv X_{0,0,\hat{p}_1,\hat{p}_1}$.

The second relation is for  the out-of-plane  polarization  and helicity transfer scatterings:
\begin{eqnarray} 
X^{pp}_{n,0,n,0}  & = & X^{pp}_{0,n,n,0} = {1\over \sigma_0} Re\left[\phi_1^{pp,\dagger} \phi^{pp}_3 - \phi^{pp,\dagger}_2 \phi^{pp}_4\right]  \nonumber \\
X^{pp}_{\hat{p}_3,0,\hat{p_1},0} & =  & X^{pp}_{0,\hat{p}_4,\hat{p_1},0} = {1\over 2\sigma_0}\left(|\phi^{pp}_1|^2 - |\phi^{pp}_2|^2\right)\cos\theta_3,
\label{ptrans}
\end{eqnarray}
where $\theta_3 = \theta_4$ for $\theta_{cm} = 90^0$.   

And the third relation is for the triple polarization case:
\begin{equation}
X^{pp}_{n,0,\hat {p}_1,s_1} = X^{pp}_{n,0,s_1,\hat {p}_1} = {1\over \sigma_0} Im\left[\phi_1^\dagger \phi_4 - \phi_2^\dagger \phi_3\right].
\label{3pl}
\end{equation}

It follows from Eqs.(\ref{psum})  and (\ref{ptrans}) that the  polarization correlations do not vanish with an increase of energy and they could be 
quite large.   For Eq.(\ref{ptrans})  the large polarization follows from the expected dominance of the real parts 
of the amplitudes and observed  hierarchy~\cite{GR} in the hard scattering regime (for any large  and fixed $\theta_{cm}$):
\begin{equation}
|\phi_1| \ge |\phi_3|\sim |\phi_4| \gg |\phi_5| \gg |\phi_2|.
\label{hier}
\end{equation}

\subsection{Signatures of Helicity Conservation}

The onset of the  helicity conservation in QCD is associated with the two following properties of helicity amplitudes: {\em first} the 
relation of Eq.(\ref{hier}) is established  with $\phi_5$ and $\phi_2$ being negligible and {\em second} the amplitudes are predominantly real.

For the particular case of the elastic $pp$ scattering at $\theta_{cm}=90^0$ the onset of the helicity conservation will result in 
the following relations between double initial state polarization asymmetries:
\begin{equation}
X^{pp}_{0,0,n,n} \approx - X^{pp}_{0,0,s_1,s_1} \approx { 1 + X^{pp}_{0,0,\hat{p}_1,\hat{p}_1} \over 2}\approx |\phi^{pp}_3|^2 =  |\phi^{pp}_4|^2
\end{equation}
and most interestingly  between  initial state polarizations and helicity transfer asymmetries:
\begin{equation}
X^{pp}_{0,0,n,n} - X^{pp}_{0,0,\hat{p}_1,\hat{p}_1} \approx -(X^{pp}_{0,0,s_1,s_1} + X^{pp}_{0,0,\hat{p}_1,\hat{p}_1})  
 \approx  {X^{pp}_{\hat{p}_3,0,\hat{p}_1,0} + X^{pp}_{0,\hat{p}_4,\hat{p}_1,0}\over 2 \cos{\theta_3}} \approx 
{|\phi^{pp}_1|^2\over 2}.
\end{equation}

For the more general case of  hard elastic $NN$ scattering   in the helicity conservation regime 
at large  and fixed $\theta_{cm}\sim 90^0$ a  rather non-trivial relation exists:
\begin{equation}
X_{\hat{p}_3,0,\hat{p}_1,0}  \rightarrow  - X_{0,0,\hat{p}_1,\hat {p}_1}
\end{equation}
which indicates that the helicity transfer asymmetry approaches to the asymmetry of double initial helicity  scattering.

For all other observables  in the helicity conservation regime one obtains expressions  which 
are given in the  RHS parts of Eqs.(\ref{Py},\ref{IniY},\ref{IniS},\ref{IniT},\ref{Triplepol}).   It is worth mentioning that 
the measurement of the triple-polarizations in Eq.(\ref{Triplepol})  are important  for verifying that the helicity amplitudes become real,
especially in relation to  $\phi_1$, $\phi_3$ and $\phi_4$ amplitudes.

\section{Quark Counting Rules of QCD}
\label{sec5}

About 40 years ago in two seminal papers~\cite{MMT,BF1},  predictions were made that at fixed $\theta_{cm}$  the high energy 
hadronic processes  probe the constituents of hadrons in their minimal Fock component states. This observation
stimulated extensive research on hard exclusive hadronic processes.  

Within  automodelism (scale invariance) assumption  the invariant energy dependence of the  Feynman amplitude, 
$\mid \langle cd\mid M\mid ab\rangle \mid^2$,  of  hard exclusive 
$a+b\rightarrow c+d$  scattering is such that it compensates  the combined dimensionality of the Fock states of participating hadrons,
$ [N_aN_bN_cN_d]^{1\over 2} = m^{n_a+n_b+n_c+n_d-4}$, where $N_j$'s are  the normalization factors and 
$n_j$'s are the number of constituents of the Fock component of the  $j$ hadron:
\begin{equation}
|j> = N_j^{1\over 2}|n_j, constituents\rangle.
\label{Fstate}
\end{equation}

This results in the simple relation for  the energy dependence of  the differential cross sections of hard exclusive reactions:
\begin{equation}
{d\sigma ^{ab\rightarrow cd}\over dt}  = F(\theta_{cm})_{ab\rightarrow cd} s^{-(n_a+n_b+n_c+n_d - 2)},
\label{cqr}
\end{equation}
where $F(\theta_{cm})_{ab\rightarrow cd}$ is a function of $\theta_{cm}$ only.

As it was observed in Refs.~\cite{Bg1,Bg2} the automodelism assumption is  justified for local field theories with vector interaction and thus it  could be 
extended to electromagnetic processes in which case leptons and bare photons would be  counted as a one unit of  constituent.

\begin{figure}[ht]
%\vspace{-0.4cm}
\centering\includegraphics[scale=0.7]{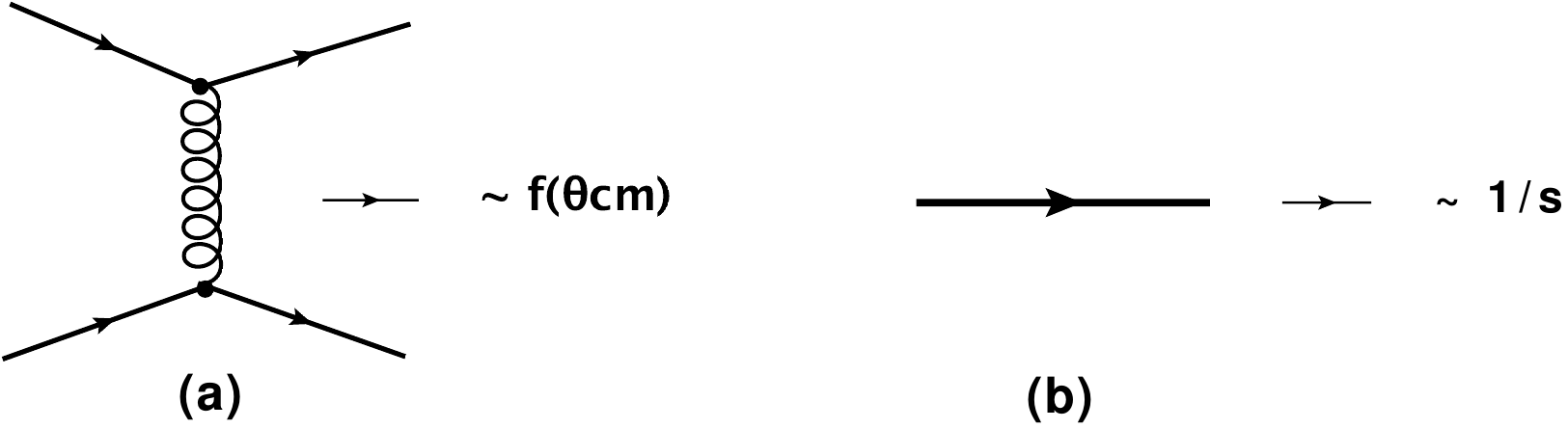}
%\vspace{-0.4cm}
\caption{Quark counting rules.}
\label{qqr}
\end{figure}

One of the most beautiful observations was~\cite{BF1,BF2} that the energy counting rule naturally follows from perturbative QCD in which the gluon exchange between 
massless quarks  will provide the scale invariance of the scattering amplitude. Within pQCD one can prove the two main asymptotic rules presented 
in Fig.\ref{qqr}:  (a) the sub-process of hard elastic $quark-quark$ interaction is energy independent (depends only on the scattering angle) and  (b) each 
energetic propagator of intermediate quark line enters with factor,  ${1\over s}$. 
These two rules in addition to the above mentioned counting of leptons and bare photons define all the necessary rules for estimating the 
energy dependence of  differential cross sections of  hard exclusive scattering of hadrons, leptons  as well as photons in the 
fixed $\theta_{cm}$ kinematics in the $s\rightarrow \infty$ limit.

The important consequence of quark-counting rules is that it explains why one should expect the dominance of the minimal-Fock component 
in the wave function of the interacting hadrons.  The simple estimate shows that higher order Fock components will result in larger negative 
exponent in the $s$-dependence of the scattering amplitude and therefore will decrease much faster than the contribution due to 
minimal-Fock component in the high energy limit.

\section{Basics of Light-Front Dynamics in QCD}
\label{sec6}

{\bf Bound State Problem:}
One of the fundamental problems in field theory is the description of the quantum  bound states 
in a relativistically  invariant form.  The problem inevitably arises  with the proper treatment of the time at which 
the bound state is observed as well as the  unambiguous  identification of the compositeness of a  bound state due to 
the non-trivial structure of the vacuum in field theories.  The latter is related to the differentiation of the constituents of the  
bound system from the particles arising from the vacuum fluctuations.

\begin{figure}[ht]
%\vspace{-0.4cm}
\centering\includegraphics[scale=1.0]{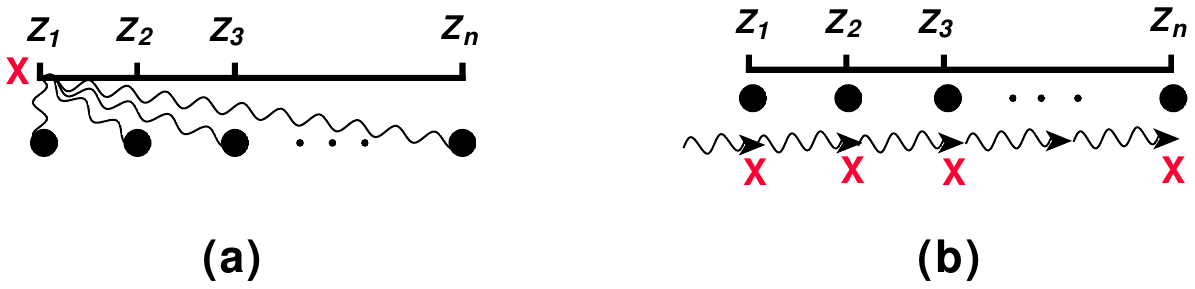}
%\vspace{-0.4cm}
\caption{Relativistic measurement of the wave function.}
\label{lf}
\end{figure}

The problem of the treatment of the time in  the relativistic dynamics of the bound states can be seen in the following 
consideration of the wave function of the system consisting of $n$ particles. To "measure" the wave function of the system, 
the observer $X$ in Fig.\ref{lf}~(a) must instantaneously measure the positions and  times  of all   $n$ particles.
In nonrelativistic dynamics due to Galilean relativity, the $n$ positions can be measured instantaneously at the given absolute time $t$,
allowing   
one to talk about the wave function of the system defined as $\Psi(z_1,z_2,z_3,É.,z_n,t)$. 

In the relativistic case,  the measurement will 
assume sending light signals from the position of the observer $X$  (Fig.\ref{lf} (a)) to all $n$ particles resulting in the measurement of the
function $\Psi(z_1,t_1,z_2,t_2,z_3,t_3,É..z_n,t_n)$.  Even if the latter function can be arranged to be relativistically invariant it does not 
have a meaning of the bound state  of the system at  the given time, $t$.

This problem can be solved if  the observer "rides" on the front   of the light beam  that passes through the all  $n$ particles.  In this case the 
observer will find that all constituents are measured at the same "light front" time  
$\tau = t_1 - z_1  = t_2 - z_2 = t_3  - z_3 \cdots t_n-z_n$ and can reconstruct the wave function of the system with constituents 
being measured at light-cone positions of ${\cal Z}_i = t_i - z_i$, $i = 1, \cdots n$  at the same light-cone time $\tau$: 
$\Psi_{LF}({\cal Z}_1,{\cal Z}_2,{\cal Z}_3,\cdots {\cal Z}_n,\tau)$.    

That the relativistic dynamics simplifies in the light-front  due to the maximum number of kinematic generators of the Lorentz group, was first observed 
by P.~A.~M.~Dirac~\cite{DiracLF}.  In relation to relativistic bound states, S.~Weinberg~\cite{SW} demonstrated how the simple structure of the vacuum 
allows  the separation of the constituent structure of the bound state from  the vacuum fluctuation contributions. 

The light-front dynamics is  a natural approach in QCD, in which due to light masses of quarks, one naturally deals with 
relativistic bound states~\cite{BrodskyLF}. In this case, choosing quantization axis by the direction of the momentum of the hadron,
its wave function is quantized at fixed $z^+ = t+z$ time with $p^-$ component representing the light-front energy of the hadron. In this 
approach the Weinberg type equation specifies the quark-gluon wave function of the hadron which depends on  the transverse momenta
 ${\bf k_{\perp,i}}$  and  the  light-front longitudinal momentum fractions  (defined as $x_i \equiv {k_i^+\over p^+}$) of the constituent quarks and gluons.
The wave function defined in  such  way is a Lorentz boost invariant and addresses the main issues  of the  description of the relativistic 
bound states  -  Lorentz invariance and definite light-cone time of the quantization.   

In practical applications,  the quark-gluon wave function of  hadrons  is expanded into 
the Fock component representations  that allows to account fully the compositeness of the hadrons as a bound system of quarks.
For example the Fock-state expansion of the light cone wave function of the nucleon can be presented as ~\cite{BroPaul,BLec2004}
\begin{eqnarray}
\mid \Psi^{S_z}_{N}(P^+,{\bf P_\perp})\rangle & =  & \sum\limits_{n}\prod\limits_{i=1}^{n}{dx_i d^2 k_{\perp,i}\over \sqrt{x_i} 16 \pi^3} 
(2\pi)^3 2\delta(1-\sum\limits_{j=1}^nx_j)\delta^{(2)}(\sum\limits_j^n{\bf k_{\perp,j}})\nonumber \\
&\times & \psi_{n/N}(x_i,{\bf k_{\perp,i}},\lambda_i)\mid n; x_i p^+, x_i{\bf P_\perp} + {\bf k_{\perp,i}}, \lambda_i\rangle 
\end{eqnarray}
where $\mid n; x_i p^+, x_i{\bf P_\perp} + {\bf k_{\perp,i}}, \lambda_i\rangle$ represents the eigenstates of  Fock-components with 
$n$ constituents.  The coefficients: $\psi_{n/N}(x_i,{\bf k_{\perp,i}},\lambda_i)$ are the  probability amplitudes (wave functions) for given $n$ constituents 
with relative  transverse momenta, ${\bf k_{\perp,i}}$ and "longitudinal" momentum fractions $x_i$.

\medskip
\medskip

\noindent {\bf Hard pQCD Interaction in the Light-Front:}
Using the light cone - helicity spinors in the form~\cite{BroLep}:
\begin{equation}
u_\lambda(p^+,p_t) = (p^+ + \beta m + \alpha_t\cdot p_t) \chi_{\lambda},
\end{equation}

where 
\begin{equation}
\chi_{\pm{1\over 2}} = {1\over \sqrt{2}} (1 + \alpha_3)\chi_{\pm},
%\end{equation}
\ \ \ \ \mbox{with} \ \ \ \
 %$ 
 \chi_+  = \left( \begin{array}{c}
1\\
0 \\
0 \\
0 \end{array} \right)
%$ 
\ \  \mbox{and} \ \ 
%$ 
\chi_-  = \left( \begin{array}{c}
0\\
1 \\
0 \\ 
0 \end{array} \right) ,
%$.
\end{equation}
for the fermion-vector field vertex one obtains:
\begin{eqnarray}
\bar u_{\lambda^\prime}(p^\prime)\gamma^+ u_{\lambda}({p})  & =  & 2\sqrt{p^\prime_+p_+} \delta^{\lambda,\lambda^\prime}\nonumber \\
\bar u_{\lambda^\prime}(p^\prime)\gamma^- u_{\lambda}({p})  & =  & {1\over \sqrt{p^\prime_+ p_+}}
\left[(1+S_{\lambda,\lambda^\prime})p_Rp^\prime_L + (1-S_{\lambda,\lambda^\prime})p_Lp^\prime_R + m^2\right]\delta^{\lambda,\lambda^\prime} \nonumber \\
 &  + & {\Delta_{\lambda,\lambda^\prime}m\over \sqrt{p^\prime_+ p_+}}\left[(p_R - p^\prime_R)(1+\Delta_{\lambda,\lambda^\prime}) + 
 (p_L - p^\prime_L)(1-\Delta_{\lambda,\lambda^\prime}\right]\nonumber \\
 \bar u_{\lambda^\prime}(p^\prime)\gamma^{R/L} u_{\lambda}({p})  & = & \left[\sqrt{{p_+\over p^\prime_{+}}}p^\prime_{R/L}(1\mp S_{\lambda\lambda^\prime})
 + \sqrt{{p^\prime_+\over p_{+}}}p_{R/L}(1\pm S_{\lambda\lambda^\prime})\right]\delta^{\lambda,\lambda^\prime}\nonumber \\
 & + & m\Delta_{\lambda,\lambda^\prime}\left[ \sqrt{{p_+\over p^\prime_{+}}} - \sqrt{{p^\prime_+\over p_{+}}}\right](1-\Delta_{\lambda,\lambda^\prime}),
\end{eqnarray}
where $"R"$  and $"L"$ indices of the given vector $A$ correspond to $A_{R/L} = A_x\pm i A_y$. Also, $S_{\lambda,\lambda^\prime} \equiv \lambda+\lambda^\prime$ and 
$\Delta_{\lambda,\lambda^\prime} \equiv  \lambda-\lambda^\prime$.

By using the light cone gauge for the exchanged gluon propagators, one avoids the necessity  of inclusion of the ghost fields to cancel the nonphysical 
contributions  of gluonic fields and, for the interaction "cell" of Fig.\ref{qqr}(a) one obtains:
\begin{equation}
\bar u\gamma^\mu u d_{\mu\nu} \bar u \gamma^\nu u = 
\bar u\left[ {\gamma^+\over r^+}r_\perp - \gamma_\perp\right]u \bar 
u \left[ {\gamma^+\over r^+}r_\perp - \gamma_\perp\right]u, 
\end{equation}
where $r$ is the four-momentum of the exchanged gluon.

Applying the above defined vertex rules, it is now easy to show,  (a) the validity of counting rules of Fig.\ref{qqr} in perturbative QCD 
and (b) that the counting rules are accompanied  by 
the helicity conservation of  interacting quarks.  Thus the signature of the quark-counting rules together with  the helicity conservation will 
represent  a powerful indicator for the onset of the pQCD regime  in strong interactions.

\subsection{Non-Perturbative QCD Contribution and Factorization Properties of  Hard Scattering Amplitudes}
Even though in probing short distance properties of $NN$ interaction our focus is on the pQCD aspects of hard 
quark-gluon interactions, we cannot avoid the non-perturbative contributions.  The latter will appear prominently  in the 
wave functions of  nucleons. The necessary assumption that allows us  to proceed with the calculation of 
hard $NN$ scattering amplitudes is the assumption of the factorization of nonperturbative and perturbative QCD contributions~\cite{BroLep}.  
This assumption is based on the very plausible expectation that the dominant contribution from the wave functions of incoming 
and outgoing nucleons  
come from the collinear states of their minimal Fock components.  In this case, each three quarks of initial nucleon "arrive" at 
hard scattering "point" with momenta almost parallel to their parent nucleons and after the hard scattering they emerge parallel 
to the respective momenta of outgoing nucleons. Clearly the wave functions of nucleons containing almost collinear quarks are 
dominated by the nonperturbative QCD contribution. 

Even if such factorization is well justified and applied,  another problem is the  number of possible diagrams (which is in the order of tens of thousands) 
which take  into account   all  possible combinations of five hard gluon exchanges  between incoming and outgoing collinear quarks. This and the 
issue of nonperturbative wave functions can be addressed if the calculated quantities are expressed  through the experimentally 
measured quantities such as nucleon-form factors ~\cite{reducedAMPs}  or  if we  consider  the observables such as 
cross section ratios, polarization and  angular distribution  asymmetries  which are not sensitive to the absolute normalization of the amplitudes and  
are sensitive either to the structure of hard scattering or  symmetry of the nucleon wave functions.

Below we will consider a few examples in which  the above approach is applied in describing the  ratios of the 
hard elastic $pn$ and $pp$ scatterings as well as the angular asymmetry of 
the hard $pn\rightarrow pn$ scattering around $\theta_{cm}=90^0$.

\section{Quark Interchange Mechanism of Hard Elastic $NN$ Scattering }
\label{sec7}

As it follows from Eq.(\ref{cqr}), the quark counting rule applied to the hard elastic $NN$ scattering predicts
\begin{equation}
{d\sigma \over dt}  = F(\theta_{cm})_{NN\rightarrow NN} \cdot s^{-10},
\end{equation}
which, as it follows from the discussion of  Sec.\ref{sec3},  describes the bulk of the energy dependence of hard $NN$ scattering 
at large $\theta_{cm}$. According to the application of the quark counting rules, the  $s^{-10}$ energy dependence requires the hard  exchange 
of  five gluons between six quarks representing the minimal Fock components of two nucleons.  
One such mechanism that satisfies the  above requirement is the quark-interchange mechanism of  the $NN$ interactions.

\subsection{Dominance of Quark-Interchange Diagrams}
\begin{figure}[ht]
%\vspace{-0.4cm}
\centering\includegraphics[scale=0.8]{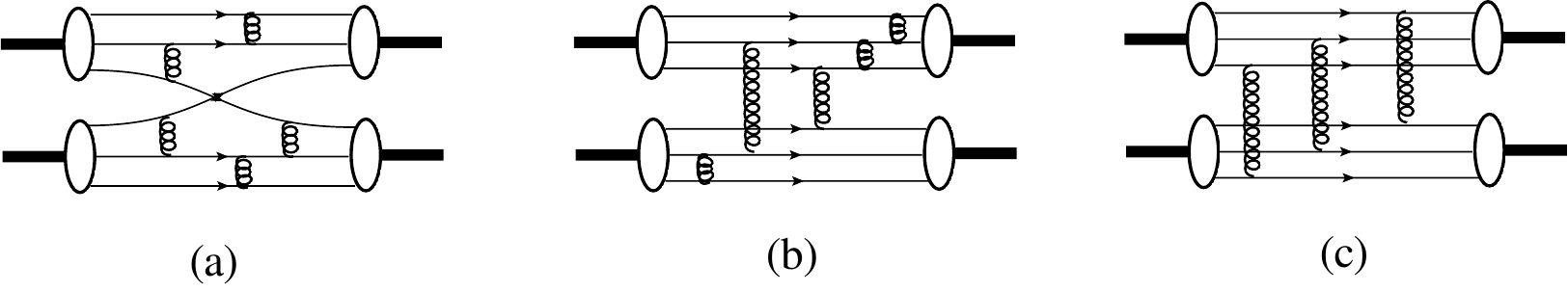}
\vspace{-0.4cm}
\caption{Possible scattering diagrams for elastic  NN processes.}
\label{QCD_NN_diagrams}
\end{figure}

Considering the topology of hard QCD scatterings in elastic NN processes one can identify a few possibilities  in which 
the collinear quarks from minimal Fock components of the nucleon wave function  can scatter into the final state (Fig.\ref{QCD_NN_diagrams}).
In fact the mechanism of independent three-gluon exchanges (Landshoff mechanism~\cite{Landshoff}) (Fig.\ref{QCD_NN_diagrams}~({c}) predicts 
softer ($s^{-8}$) energy dependence since it requires less number of gluon exchanges. However this diagram 
(as well as the one in Fig.\ref{QCD_NN_diagrams}(b)) does not exchange  quarks and will contribute equally to $NN$ and $\bar N N$ scattering.
Thus, comparison of the experimental data of   these two reactions  will allow the estimation of the contributions of such diagrams. 
Such a comparison was made in the experiment  of Ref.~\cite{h20}  which found the upper limit of the ratio of elastic differential cross sections of 
$\bar p p$ to $pp$ at $p_{Lab}=9.9$~GeV/c at $\theta_{cm}=90^0$ to be less than $4\%$.   The measurement of the same ratio at 
$p_{Lab}=5.9$~GeV/c yielded $\sim 2.5\%$.   These results, together with the comparison  of  
the cross sections of twenty different hard exclusive hadronic reactions  unambiguously confirmed the dominance of    
the hard  scattering amplitudes  of  hadrons  containing quarks  of  the same  flavor with respect to the 
scattering amplitudes of hadrons that share no common flavors of quarks. 

Thus, one can start with a rather solid experimental justification of the dominance of the quark-interchange (QIM) 
diagrams of Fig.\ref{QCD_NN_diagrams}(a) 
in the hard elastic NN scattering amplitude. In fact the theoretical analysis demonstrates that the QIM diagrams 
 represent the dominant mechanism of hard elastic scattering for up to ISR energies  (see discussion in ~\cite{BLC}). 
 It is worth mentioning that  the quark-interchange mechanism also describes reasonably  well 
 the $90$ c.m. hard  break-up  of two nucleons from the deuteron~\cite{gdpn,gdpnpol,gdpnpp,gdpn2} and $^3He$~\cite{hrm}.
 
\begin{figure}[ht]
\centering\includegraphics[height=3cm,width=8cm]{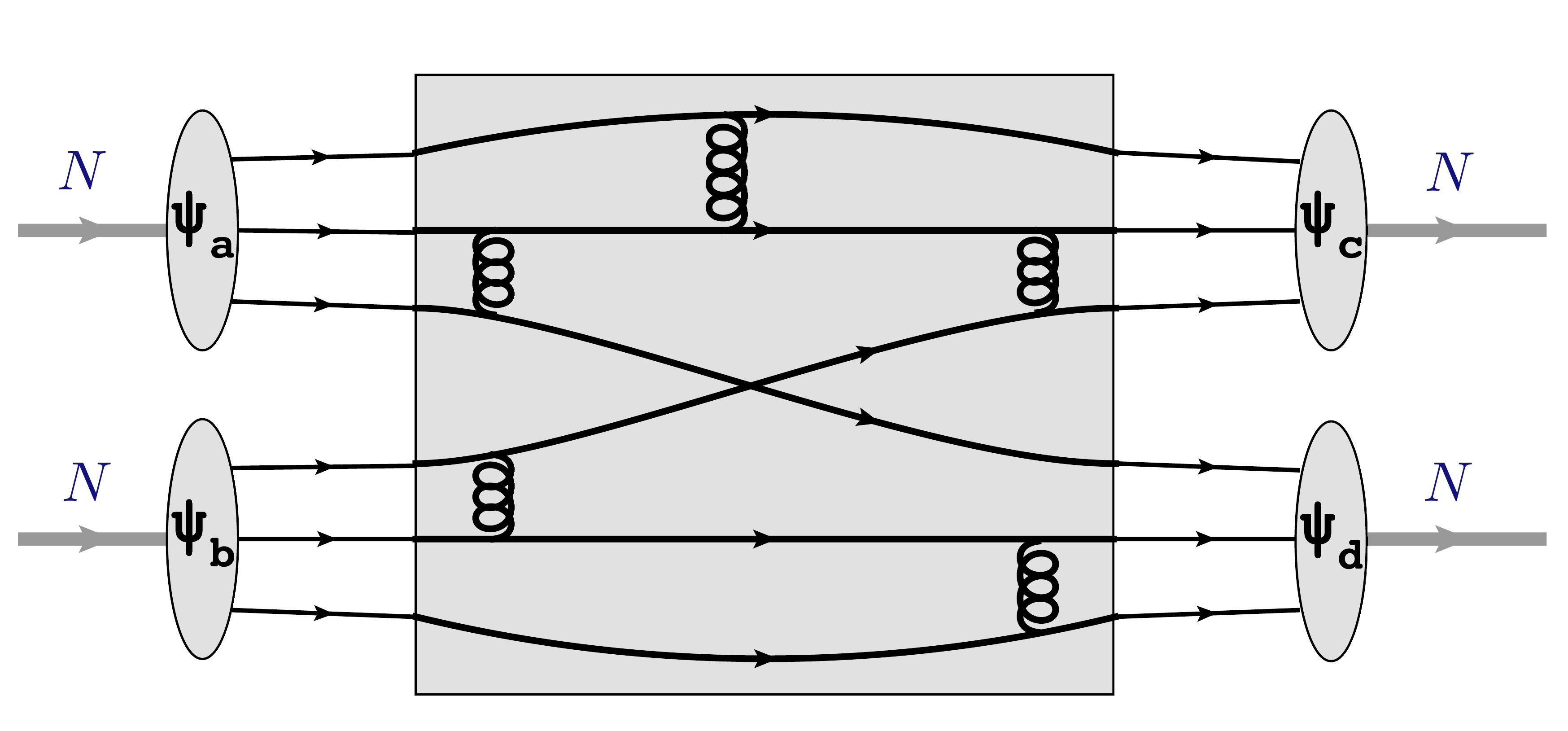}
\vspace{-0.2cm}
\caption{Typical diagram for quark-interchange mechanism of $NN\rightarrow NN$ scattering.}
\vspace{-0.32cm}
\label{QIM}
\end{figure}

Based on the above arguments we now focus on the quark-interchange diagrams which contain  
five gluon exchanges as shown in Fig.\ref{QIM}.    
Furthermore, we  assume the validity of earlier  discussed  factorization of the minimal (3q) Fock components of 
initial and final state wave functions of nucleons 
and the hard  interaction kernel,  which represents the quark interchange  scattering through the five hard gluon exchanges. 
These assumption allows to represent 
the amplitude of  hard  $N(a)+N(b)\rightarrow N({c})+N(d)$ scattering of Fig.\ref{QIM}, 
within quark-interchange approximation  in the following form~\cite{pnangle}:
\begin{eqnarray}
& & \langle cd\mid T\mid ab\rangle  = 
\sum\limits_{\alpha,\beta,\gamma} 
\langle  \psi^\dagger_c\mid\alpha_2^\prime,\beta_1^\prime,\gamma_1^\prime\rangle
\langle  \psi^\dagger_d\mid\alpha_1^\prime,\beta_2^\prime,\gamma_2^\prime\rangle \times
\nonumber \\
& &  \ \ \ \ 
\langle \alpha_2^\prime,\beta_2^\prime,\gamma_2^\prime,\alpha_1^\prime\beta_1^\prime
\gamma_1^\prime\mid H\mid
\alpha_1,\beta_1,\gamma_1,\alpha_2\beta_2\gamma_2\rangle\cdot 
\langle\alpha_1,\beta_1,\gamma_1\mid\psi_a\rangle
\langle\alpha_2,\beta_2,\gamma_2\mid\psi_b\rangle,\nonumber \\
\label{ampl}
\end{eqnarray}
where ($\alpha_i, \alpha_i^\prime$), ($\beta_i,\beta_i^\prime$) and ($\gamma_i,\gamma_i^\prime$) 
describe the spin-flavor quark states of minimal-Fock component of nucleon wave function before and after the hard 
scattering, $H$,  and 
\begin{equation}
C^{j}_{\alpha,\beta,\gamma} \equiv \langle\alpha,\beta,\gamma\mid\psi_j\rangle
\label{Cs}
\end{equation}
describes the probability amplitude of finding the $\alpha,\beta,\gamma$ helicity-flavor 
combination of three valence quarks in the nucleon $j$~\cite{FG,pnangle}. 
Note that in Eq.(\ref{ampl}) the factorization of nucleon wave functions from 
the hard scattering kernel is justified by the energies, characteristic to  the 
$C^{j}_{\alpha,\beta,\gamma}$ factors, being
of the order of the nucleon mass,  while the $H$-kernel is characterized by  the 
transfered momenta $-t,-u\gg m_N^2$.

To be able to calculate the $C^{j}_{\alpha,\beta,\gamma}$ factors, one  
represents the nucleon wave function through the helicity-flavor basis of the valence  
quarks.  We use a rather general form, separating  the nucleon wave function into two parts characterized 
by two (e.g. second and third)  quarks being  in spin zero - isosinglet and spin one - isotriplet states 
as follows:
\begin{eqnarray}
\psi^{i^3_{N},h_N} & & = {1\over \sqrt{2}}\left\{
\Phi_{0,0}(k_1,k_2,k_3)
(\chi_{0,0}^{(23)}\chi_{{1\over2},h_N}^{(1)})\cdot
(\tau_{0,0}^{(23)}\tau_{{1\over 2},i_N^{3}}^{(1)}) 
\right.  +  \Phi_{1,1}(k_1,k_2,k_3) 
 \times \nonumber \\
& & 
\sum\limits_{i_{23}^3=-1}^{1}  \sum\limits_{h_{23}^3=-1}^{1}
\langle 1,h_{23}; {1\over 2},h_{N}-h_{23}\mid {1\over 2},h_N\rangle
\langle 1,i^3_{23}; {1\over 2},i^3_{N}-i^3_{23}\mid {1\over 2},i^3_N\rangle
\times\nonumber \\
& & 
\ \ \ \ \ \ \ \  \ \ \ \ \ \ \ \ \ \  \ \ \ \  \left. (\chi_{1,h_{23}}^{(23)}\chi_{{1\over2},h_N-h_{23}}^{(1)})\cdot
(\tau_{1,i^3_{23}}^{(23)}\tau_{{1\over 2},i_N^{3}-i^3_{23}}^{(1)})\right\},
\label{wf}
\end{eqnarray}
where $j_N^3$ and $h_N$ are the isospin component  and the helicity of the nucleon.
Here  $k_i$'s are the light cone momenta of quarks which should be understood as  
($x_i,k_{i\perp}$), where $x_i$ is a light cone momentum fraction of the nucleon 
carried by the $i$-quark. 
We define $\chi_{j,h}$ and $\tau_{I,i^3}$ as helicity 
and isospin  wave functions, where $j$ is the spin, $h$ is the helicity, 
$I$ is the isospin and $i^3$ its third component.
The Clebsch-Gordan coefficients are 
defined as $\langle j_1,m_1;j_2,m_2\mid j,m\rangle$. 
Here, $\Phi_{I,J}$ represents the momentum dependent part of the wave function 
for ($I=0,J=0$) and ($I=1,J=1$) two-quark spectator states respectively.
We also introduce  a  parameter, $\rho$: 
\begin{equation}
\rho = {\langle \Phi_{1,1}\rangle \over \langle \Phi_{0,0}\rangle },
\label{rho}
\end{equation}
which characterizes an average relative  magnitude of 
the wave function components corresponding to 
($I=0,J=0$) and ($I=1,J=1$) quantum  numbers  of two-quark ``spectator'' states.
Note that the two extreme values of $\rho$ define two well known approximations:
$\rho=1$ corresponds to the exact SU(6) symmetric picture of the nucleon 
wave function  and 
$\rho=0$ will correspond to the contribution of only good-scalar diquark configuration 
in the nucleon wave function (see e.g. Ref.~\cite{Anselmino,RJ,SW} where this component is referred to as 
a scalar or good diquark configuration~($[qq]$) as opposed to a vector or bad diquark 
configuration denoted by  $(qq)$).  
In further discussions we will keep $\rho$ as a free parameter. 
Note that in our approach diqaurks represent a  
$qq$-component of the nucleon wave function and  quarks from the diquark 
participate in the hard scattering through the quark-interchange.

To calculate the scattering amplitude of Eq.(\ref{ampl}) we take into account the  
conservation of the helicities of quarks participating in the hard scattering. 
This allows us to approximate the hard scattering part of the amplitude, $H$,  
in the following form:
\begin{equation}
H \approx \delta_{\alpha_1\alpha_1^\prime}\delta_{\alpha_2\alpha_2^\prime}
\delta_{\beta_1,\beta_1\prime}
\delta_{\gamma_1,\gamma_1^\prime}
\delta_{\beta_2,\beta_2\prime}
\delta_{\gamma_2,\gamma_2^\prime} {F(\theta)\over s^4}.
\label{H}
\end{equation}
Inserting this expression into Eq.(\ref{ampl}) for the QIM 
amplitude one obtains~\cite{FG,pnangle}:
\begin{equation}
\langle cd\mid T\mid ab\rangle = Tr(M^{ac}M^{bd})
\label{ampl2}
\end{equation}
with:
\begin{equation}
M^{i,j}_{\alpha,\alpha^\prime} = 
C^{i}_{\alpha,\beta\gamma}C^{j}_{\alpha^\prime,\beta\gamma} + 
C^{i}_{\beta\alpha,\beta}C^{j}_{\beta\alpha^\prime,\beta} + 
C^{i}_{\beta\gamma\alpha}C^{j}_{\beta\gamma\alpha^\prime},
\label{QIMMs}
\end{equation} 
where we sum over all the  possible values of $\beta$ and $\gamma$.  

Furthermore, separating the energy dependence and angular parts  for nonvanishing 
 helicity amplitudes of Eq.(\ref{phis}) one obtains:\\
for $pp\rightarrow pp$:
\begin{eqnarray}
\phi_1 & = &    C(s)\left[(3 + y)F(\theta)    +  (3 + y)  F(\pi-\theta)\right] \nonumber \\
\phi_3 & = &    C(s)\left[(2 - y)F(\theta)    +  (1 + 2y) F(\pi-\theta)\right] \nonumber  \\
\phi_4 & = &   -C(s)\left[(1 + 2y)F(\theta)  +  (2 - y)  F(\pi-\theta)\right]
\label{pppp}
\end{eqnarray}
and for  $pn\rightarrow pn$:
\begin{eqnarray}
\phi_1 & = &  C(s)\left[(2 - y)F(\theta)   +  (1 + 2y)  F(\pi-\theta)\right] \nonumber \\
\phi_3 & = &  C(s)\left[(2 + y)F(\theta)   +  (1 + 4y) F(\pi-\theta)\right] \nonumber \\
\phi_4 & = &   C(s)\left[2y F(\theta)      +     2y  F(\pi-\theta)\right]
\label{pnpn}
\end{eqnarray}
with $\phi_2=\phi_5=0$ due to helicity conservation. Here the function $C(s)={N\over s^{4}}$ is due to quark-counting rule with 
the normalization factor, N,  and $F(\theta)$ accounts for the angular dependence generated by hard scattering kernel. 
In the above equation the parameter:
\begin{equation}
y = x(x+1) \ \ \mbox{with } x = {2\rho \over 3(1+\rho^2)} 
\label{xy}
\end{equation}
defines the symmetry of the minimal-Fock (3q) component of nucleon wave function. 
For example,  $\rho=1$ case reproduces the SU(6) result of Refs.~\cite{FG} and ~\cite{BLC}.

As Eqs.(\ref{pppp}) and (\ref{pnpn}) show the helicity amplitudes reveal strong sensitivity to the underlaying symmetry 
of $3q$ nucleon wave function.  The hard kernel, $H$, as it follows from Eq.(\ref{H}) defines mainly the overall $s$- dependence 
of the amplitudes, thus  verifying the   
dominance of  the minimal-Fock component of the quark wave function of 
nucleon and the QCD mechanism of  hard interaction.  What concerns  the overall normalization of the helicity amplitudes they 
are defined both by the normalization of the nucleon wave function and the sum of the multitude of  hard scattering 
amplitudes.

In the following we will consider three examples in which the specific measurements can verify the  different aspects of 
the above described model of  NN interaction.\\

\noindent{\bf Ratio of Differential Cross Sections of Elastic $pn$ and $pp$ Scatterings at \boldmath{$\theta_{cm}=90^0$}:}
As it can be checked  from Eqs.(\ref{pppp}) and (\ref{pnpn}), at $\theta_{cm}=90^0$  the helicity amplitudes of 
$pp$ scattering depend identically on the helicity-flavor symmetry factor, $y$   of the $3q$ wave function of the proton, while 
no such identity exists for  $pn$ amplitudes.   Therefore one should expect rather strong sensitivity  
of the ratio of $pn$ to $pp$ elastic cross sections on the underlying spin-flavor symmetry of nucleon wave functions~\cite{gdbb}.
\begin{figure}[ht]
%\vspace{-0.4cm}
\centering\includegraphics[height=7cm,width=9.3cm]{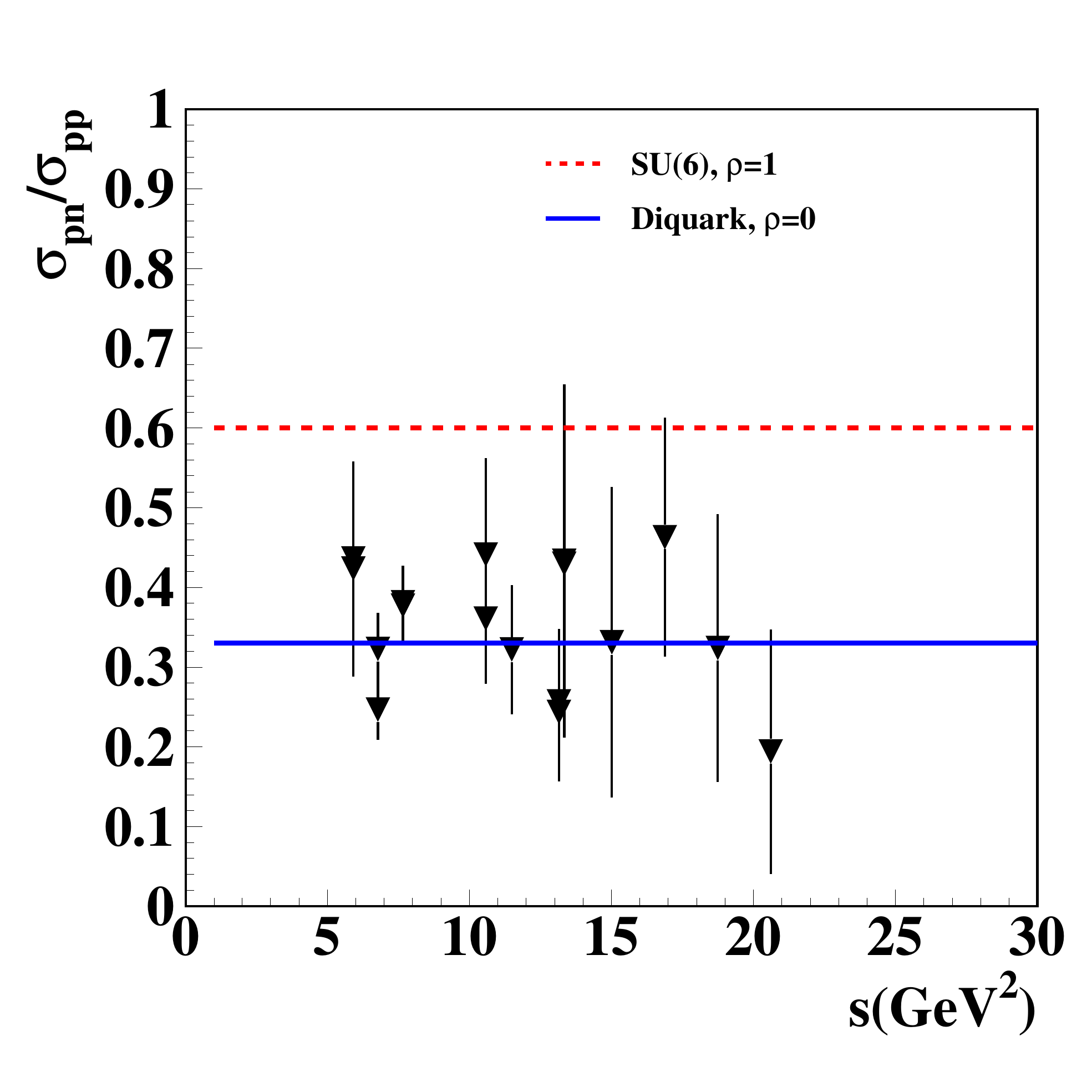}
%\vspace{-0.4cm}
\caption{Ratio of the  $pn\rightarrow pn$ to $pp\rightarrow pp$   elastic differential cross sections 
as a function of $s$ at $\theta^{N}_{c.m.}=90^0$. The data are from Refs.~\cite{Allaby,Akerlof,pnexp1,pnexp2}.}
\label{pn_pp_ratio}
\end{figure}
These can be seen in Fig.\ref{pn_pp_ratio}, where the ratio of the differential cross sections of elastic $pn$ to $pp$ scatterings  
is  presented as a function of $s$ at $\theta_{cm}=90^0$.  The calculations are performed for the case of $SU(6)$ symmetry ($\rho=1$) and 
scalar diquark model, for which  $\rho=0$.  The results are compared with the available experimental data.  
Despite the large errors present in the data,  one can clearly see the sensitivity of the considered ratio to the symmetry structure of nucleon wave function with data favoring the  scalar diquark model of nucleon.\\

\noindent{\bf Angular Asymmetry  of Elastic $pn$ Scattering:}
In this  example we demonstrate that an  observable such as the asymmetry of 
a hard elastic proton-neutron scattering with respect to $\theta_{cm}=90^0$   may 
provide another  insight into the  helicity-flavor symmetry of 
the 3q wave function  of the nucleon. The measurable quantity we consider is:
\begin{equation}
A_{90^0}(\theta) = 
{\sigma(\theta) - \sigma(\pi -\theta)\over \sigma(\theta) + \sigma(\pi-\theta)},
\label{Asym}
\end{equation}
where $\sigma(\theta)$  is the differential cross section of the elastic 
$pn$ scattering. 

\begin{figure}[t]
\centering\includegraphics[height=7cm,width=10cm]{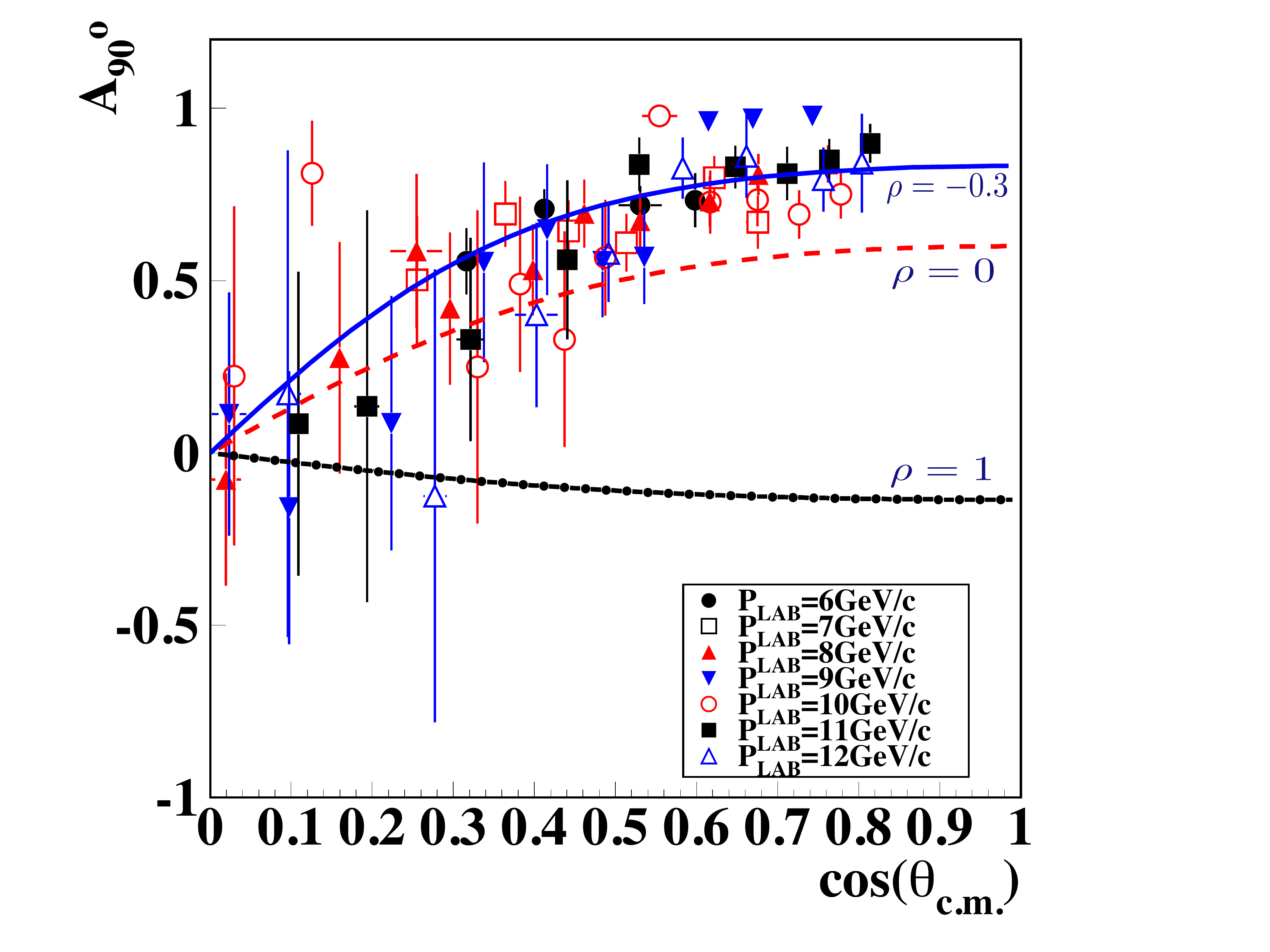}
\vspace{-0.1cm}
\caption{Asymmetry of $pn$ elastic cross section. Solid dotted line - SU(6), with $\rho=1$,
dashed line - diquark model with $\rho=0$, solid line - fit with $\rho=-0.3$. The data are from Refs.~\cite{pnexp1,pnexp2}.}
\label{ang_asym}
\end{figure}

In Fig.\ref{ang_asym} the asymmetries 
of $pn$ scattering calculated with SU(6)~($\rho=1$) and pure scalar-diquark~($\rho=0$) models are 
compared with the data~\cite{pnangle}.  
%In these estimates we use 
%$F(\theta) = C\cdot sin^{-2}(\theta)(1-cos(\theta))^{-2}$ 
%dependence of the angular function~\cite{RS} which is consistent with the 
%picture of hard collinear QIM scattering of valence quarks with five gluon exchanges and  
%reasonably well reproduces the main characteristics  of the  angular dependencies  of 
%both $pp$ and $pn$ elastic scatterings. Note that using a form of the angular function 
%based on nucleon form-factor arguments~\cite{BCL,FGST}, $F\approx (1-cos(\theta))^{-2}$
%will result in the same angular asymmetry.
The comparisons show that the nucleon wave function~(\ref{wf}) with the good-scalar 
diquark component~($\rho=0$) produces the right sign for the angular asymmetry.
On the other the data are in qualitative disagreement  with the prediction of
exact SU(6) symmetry~($\rho=1$) of the quark wave 
function of nucleon.

In Ref.~\cite{pnangle} we used $\rho$ as a free parameter to obtain the best fit of the data, which 
was found for 
\begin{equation}
\rho \approx  -0.3\pm 0.2.
\label{rhofit}
\end{equation}
The nonzero magnitude of $\rho$ indicates  the small but finite relative strength of 
a bad/vector diquark  configuration in the nucleon wave function.   
It is intriguing that the obtained magnitude of $\rho$ is consistent with  the $10\%$ 
probability of ``bad'' diquark configuration discussed in Ref.~\cite{SW2}.

Another interesting property of Eq.(\ref{rhofit}) is the negative sign of the  
parameter $\rho$.
Within  qualitative  quantum-mechanical picture, the 
negative sign of $\rho$  may indicate for example the 
existence of a repulsion in the quark-(vector- diquark) channel as opposed to 
the attraction in the quark - (scalar-diquark) channel.  It is rather surprising that 
both the magnitude and sign agree with the result of the phenomenological interaction 
derived in the one-gluon exchange quark model discussed in Ref.~\cite{RJ}.\\

\noindent{\bf The \boldmath{$A_{nn}$} Asymmetry  of Elastic $pn$ Scattering:}
In this last example we consider the $A_{nn}\equiv X_{0,0,n,n}$ asymmetry defined according to Eq.(\ref{IniY}). 
This asymmetry is especially interesting to consider for $\theta_{cm}=90^0$ scattering, since in this case 
as it was mentioned before the all non-zero helicity amplitudes depend identically on the helicity-flavor factor of $y$.
As a result, by using helicity amplitudes of Eq.(\ref{pppp}) in  the RHS part of Eq.(\ref{IniY}) one obtains a 
constant value,  $A_{nn}=0.333$,  at  $\theta_{cm}=90^0$ independent on the specific helicity-flavor symmetry of 3q wave functions. 

This result is compared with the data in Figs.\ref{ann} and \ref{ct_ann_pp}, which demonstrates that agreement 
with the data happens at $s\approx 18-20$~GeV$^2$ at the very same energy range where transparency studies demonstrate 
the drop of  the absorption  of propagating 
"protons" in the nuclear medium (Fig.(\ref{ct}) and (\ref{ct_ann_pp})).

\section{Conclusions and Outlook}
\label{sec8}
In this work we reviewed the current status of QCD studies of hard $NN$ processes that probe quark-gluon interactions at short 
distances.  We gave  brief history of these studies and highlighted the outstanding questions that may 
be resolved in the future experiments at the emerging facilities worldwide. We outlined the theoretical foundation of QCD hard 
processes and demonstrated that in high energy regime the hard processes probe the minimal (3q) Fock component of 
the nucleon wave function which is factorized from the  kernel that defines the dynamics of the hard scattering.

We considered few  examples of application  of the  outlined approach and demonstrated that even with the limited 
quantity and quality of the data one can conclude that most probably SU(6) symmetry is breaking down at large $x$ 
limit of the valence quark component of nucleon wave function.  Another observation we made is that, apparently, the phenomenology of 
hard $pp$ and $pn$ scattering favors the diquark picture of   the nucleon with small and finite mixture from 
the vector diquarks which enter with the negative phase relative to the scalar diquark contribution.   Finally, our 
estimate of the $A_{nn}$ asymmetry of elastic $pp$ scattering within pQCD indicates that it agrees with the data at 
same energies at which color transparency signature is observed in  the  hard $A(p,2p)X$ reactions.
 
This review covered only the hard elastic NN scatterings. However, it is worth noting  that there is a growing activity in studies of 
NN  bound systems at short distances by probing short range correlations in the nuclear wave functions 
(see e.g. ~\cite{srcrev,FSDM,srcprogress,edepn,edepnexp}).  
Currently these  studies unambiguously identified the tensor component of short-range $NN$ interactions in the nuclear medium.  
The planned  experiments at the 12~GeV energy upgraded Jefferson Lab~\cite{hnm,JLab12} will  be able to probe  the bound $NN$ systems at distances relevant  to the nuclear core, where  one may expect the onset of QCD degrees of freedom in the 
similar way as in the hard NN interactions.

Overall new experiments in studies of  both hard $NN$ scattering processes and $NN$ short-range correlations  in nuclei will provide the 
necessary ground for  advancing the understanding of QCD dynamics of  strong forces at short distances.

\medskip
\medskip

The author is  thankful to Drs.~W.~Boeglin, S.~J.~Brodsky,~R.~Gilman, L.~Frankfurt,~G.~Miller, E.~Piasetzky and M.~Strikman for  
illuminating discussions and many years of collaboration on  the physics of hard processes. Special thanks to 
Dr.  Igor Strakovsky for the suggestion and opportunity to write this review.
This work is supported by the  U.S. DOE  grant under contract DE-FG02-01ER41172.

%++ Other useful packages to be used when needed ----------------------------------------------------
%\usepackage{epsfig}
%\usepackage{amssymb}
%\usepackage{amsmath}
%\usepackage{amsfonts}
%\usepackage{amsthm,amscd}
%\usepackage{amsbsy}
%\usepackage{latexsym}
%\usepackage{bm}
%\usepackage{url} 			%% Nicely format and linebreak URLs in the bibliography (and elsewhere).
%\usepackage{layout}
%\usepackage{pslatex}
%\usepackage{cite}
%\usepackage{fleqn} 		% displayed formulas flush left (default is centered).
%\usepackage{makeidx}
%\makeindex 						% Creates index at the end of the book (with makeidx).
%\usepackage{layout} 	% To see the current values of these dimensions, use the layout package, 
											% which defines a \layout command.
%\usepackage{epstopdf}	% Automatically converts EPS files to encapsulated PDF files (using ghostscript).
%\usepackage{color}
%\usepackage{hyperref}
%\usepackage[latin1]{inputenc}
%\usepackage[T1]{fontenc}
%\raggedbottom                         %%% do NOT increase spaces between
                                      %%% paragraphs in order to fill always
                                      %%% the whole page
%\usepackage{poligraf} %% Color separation
%\usepackage[letter,cam,center]{crop} %% Printing crop-marks
%\usepackage{type1cm} 	%% Use scalable, PostScript Type 1 versions of the Computer Modern fonts.
%\usepackage{courier}	%% Replace the standard Computer Modern Typewriter font LaTeX uses
										 	%% for monospace text with the PostScript font Adobe Courier.
%\usepackage{lscape} 	% for landscape section
%\input tcilatex 			%% for Sci Word files
%++------------------------------------------------------------------------------

%\label{lastpage-01}


\begin{thebibliography}{13}
\bibitem{Rutherford1}  E.~Rutherford,
  %``The scattering of alpha and beta particles by matter and the structure of the atom,''
  Phil.\ Mag.\  {\bf 21}, 669-688 (1911).

\bibitem{Rutherford2} E.~Rutherford, Phil. \ Mag. \ {\bf 37}, 571-580 (1919).
\bibitem{Rutherford3}E.~Rutherford, Phil. \  Mag. \ {\bf 37}, 581-587 (1919).
\bibitem{BW} John M. Blatt and Victor F. Weisskopf, {\em Theoretical Nuclear Physics}, Dover Books on Physics,  Dover Publications, 
September, 2010.

\bibitem{Jastrow} R.~Jastrow, 
%{\em On the Nucleon Nucleon Interactions},
Phys. Rev. {\bf 81}, 165-170 (1951).

\bibitem{threereg}    M.~Taketani, S.~Nakamura and M.~Sasaki, Proc. Theor, Phys. (Kyoto) {\bf 6}, 581-586 (1951).
  %``The Meson theory of nuclear forces and nuclear structure,''
%  Adv.\ Nucl.\ Phys.\  {\bf 19}, 189 (1989).
   

\bibitem{EFT1} 
  D.~B.~Kaplan, M.~J.~Savage and M.~B.~Wise,
  %``Two nucleon systems from effective field theory,''
  Nucl.\ Phys.\ B {\bf 534}, 329-355 (1998).
  
 \bibitem{EFT2}  E.~Epelbaum,
  %``Nuclear Forces from Chiral Effective Field Theory: A Primer,''
  arXiv:1001.3229 [nucl-th];
 \bibitem{EFT3}   E.~Epelbaum,
  %``Few-nucleon forces and systems in chiral effective field theory,''
  Prog.\ Part.\ Nucl.\ Phys.\  {\bf 57}, 654 (2006).
 %~\cite{Yang:2009pn}
\bibitem{EFT4} 
  C.~-J.~Yang, C.~.Elster and D.~R.~Phillips,
  %``Subtractive renormalization of the NN interaction in chiral effective theory up to next-to-next-to-leading order: S waves,''
  Phys.\ Rev.\ C {\bf 80}, 044002 (2009).    
 
 \bibitem{OBEP}    R.~Machleidt,
  %``The Meson theory of nuclear forces and nuclear structure,''
  Adv.\ Nucl.\ Phys.\  {\bf 19}, 189 (1989).
 
\bibitem{NNpars}R.~B.~Wiringa, V.~G.~J.~Stoks and R.~Schiavilla,
  %``An Accurate nucleon-nucleon potential with charge independence breaking,''
  Phys.\ Rev.\ C {\bf 51}, 38 (1995).
 
 
\bibitem{SAID} 
  R.~A.~Arndt, I.~I.~Strakovsky and R.~L.~Workman,
  %``Nucleon nucleon elastic scattering to 3 GeV,''
  Phys.\ Rev.\ C {\bf 62}, 034005 (2000).
  
  \bibitem{SAIDupdt} 
  R.~A.~Arndt, W.~J.~Briscoe, I.~I.~Strakovsky and R.~L.~Workman,
  %``Updated analysis of NN elastic scattering to 3-GeV,''
  Phys.\ Rev.\ C {\bf 76}, 025209 (2007).

\bibitem{Nijmegen} 
  J.~R.~Bergervoet, P.~C.~van Campen, R.~A.~M.~Klomp, J.~L.~de Kok, T.~A.~Rijken, V.~G.~J.~Stoks and J.~J.~de Swart,
  %``Phase shift analysis of all proton proton scattering data below T(lab) = 350-MeV,''
  Phys.\ Rev.\ C {\bf 41}, 1435 (1990).


  
\bibitem{Feynman}  R.~P.~Feynman, ``Photon-hadron interactions,''
  Advanced Book Classics, Addison-Wesley Reading, Massachusetts 282p, 1998.


\bibitem{SaturneII}
\bibitem{Gaillard:1991iw} 
  G.~Gaillard, P.~Bach, J.~Ball, R.~Binz, J.~Bystricky, P.~Demierre, J.~M.~Fontaine and J.~P.~Goudour {\it et al.},
  %``The Nucleon-nucleon program at Saturne-II,''
  Helv.\ Phys.\ Acta {\bf 64}, 201 (1991).


\bibitem{J-PARC}   S.~Kumano,
  %``Hadron physics at J-PARC,''
  Nucl.\ Phys.\ A {\bf 782}, 442 (2007).
   
\bibitem{PANDA} 
  M.~F.~M.~Lutz {\it et al.}  [PANDA Collaboration],
  ``Physics Performance Report for PANDA: Strong Interaction Studies with Antiprotons,''
  arXiv:0903.3905 [hep-ex].


\bibitem{NICA} 
  A.~N.~Sissakian {\it et al.}  [NICA Collaboration],
  %``The nuclotron-based ion collider facility (NICA) at JINR: New prospects for heavy ion collisions and spin physics,''
  J.\ Phys.\ G {\bf 36}, 064069 (2009).
  
  
\bibitem{HIAF} 
  J.~C.~Yang, J.~W.~Xia, G.~Q.~Xiao, H.~S.~Xu, H.~W.~Zhao, X.~H.~Zhou, X.~W.~Ma and Y.~He {\it et al.},
  %``High Intensity heavy ion Accelerator Facility (HIAF) in China,''
  Nucl.\ Instrum.\ Meth.\ B {\bf 317}, 263 (2013).


    
\bibitem{BBL73} J.~F.~Gunion, S.~J.~Brodsky and R.~Blankenbecler,
  %``Large Angle Scattering and the Interchange Force,''
  Phys.\ Rev.\ D {\bf 8}, 287 (1973).
\bibitem{BBL75}D.~W.~Sivers, S.~J.~Brodsky and R.~Blankenbecler,
  %``Large Transverse Momentum Processes,''
  Phys.\ Rept.\  {\bf 23}, 1 (1976).

\bibitem{Landshoff}    P.~V.~Landshoff,
  %``Model for elastic scattering at wide angle,''
  Phys.\ Rev.\ D {\bf 10}, 1024 (1974).
  %%CITATION = PHRVA,D10,1024;%%
  %258 citations counted in INSPIRE as of 31 Jan 2014


\bibitem{MMT} V.~A.~Matveev, R.~M.~Muradian and A.~N.~Tavkhelidze,
  %``Automodellism in the large - angle elastic scattering and structure of hadrons,''
  Lett.\ Nuovo Cim.\  {\bf 7}, 719 (1973).

\bibitem{BF1}  S.~J.~Brodsky and G.~R.~Farrar,
  %``Scaling Laws at Large Transverse Momentum,''
  Phys.\ Rev.\ Lett.\  {\bf 31}, 1153 (1973).
 \bibitem{BF2}
  S.~J.~Brodsky and G.~R.~Farrar,
  %``Scaling Laws for Large Momentum Transfer Processes,''
  Phys.\ Rev.\ D {\bf 11}, 1309 (1975).
 
 \bibitem{FG}G.~R.~Farrar, S.~A.~Gottlieb, D.~W.~Sivers and G.~H.~Thomas,
  %``Constituent Description of n n Elastic Scattering Observables at Large Angles,''
  Phys.\ Rev.\ D {\bf 20}, 202 (1979).
 

\bibitem{BLC}   S.~J.~Brodsky, C.~E.~Carlson and H.~J.~Lipkin,
  %``Spin Effects in Large Transverse Momentum Exclusive Scattering Processes,''
  Phys.\ Rev.\ D {\bf 20}, 2278 (1979).


\bibitem{xpol1}I.~P.~Auer, D.~Hill, R.~C.~Miller, K.~Nield, B.~Sandler, Y.~Watanabe, A.~Yokosawa and A.~Beretvas {\it et al.},
  %``Measurement of the Spin Spin Correlation Parameter C(SS) in p p Elastic Scattering at 6-GeV/c,''
  Phys.\ Rev.\ Lett.\  {\bf 37}, 1727 (1976).

\bibitem{xpol2}I.~P.~Auer, A.~Beretvas, E.~Colton, D.~Hill, K.~Nield, H.~Spinka, D.~Underwood and Y.~Watanabe {\it et al.},
  %``Measurements of the Total Cross-Section Difference and the Parameter C(LL) in p p Scattering with Longitudinally Polarized Beam and Target,''
  Phys.\ Lett.\ B {\bf 70}, 475 (1977).

\bibitem{Crabb} 
   D.~G.~Crabb, R.~C.~Fernow, P.~H.~Hansen, A.~D.~Krisch, A.~J.~Salthouse, B.~Sandler, K.~M.~Terwilliger and J.~R.~O'Fallon {\it et al.},
  %``Spin Dependence of High p-Transverse**2 Elastic p p Scattering,''
  Phys.\ Rev.\ Lett.\  {\bf 41}, 1257 (1978).

\bibitem{Crosbie} 
  E.~A.~Crosbie, L.~G.~Ratner, P.~F.~Schultz, J.~R.~O'Fallon, D.~G.~Crabb, R.~C.~Fernow, P.~H.~Hansen and A.~D.~Krisch {\it et al.},
  %``Energy Dependence of Spin Spin Effects in p p Elastic Scattering at 90-Degrees Center-Of-Mass,''
  Phys.\ Rev.\ D {\bf 23}, 600 (1981).
 
  
 \bibitem{Hoyer} P.~Hoyer,
  %``Bound states -- from QED to QCD,''
  arXiv:1402.5005 [hep-ph].
  

 \bibitem{Isgur} 
  K.~Maltman and N.~Isgur,
  %``Nuclear Physics and the Quark Model: Six Quarks with Chromodynamics,''
  Phys.\ Rev.\ D {\bf 29}, 952 (1984).
 
  \bibitem{Harvey}  M.~Harvey,
  %``Effective nuclear forces in the quark model with Delta and hidden color channel coupling,''
  Nucl.\ Phys.\ A {\bf 352}, 326 (1981).

\bibitem{Obukhovsky:1982ci} 
  I.~T.~Obukhovsky, Y.~.F.~Smirnov and Y.~.M.~Chuvilsky,
  %``On The Construction Of Wave Functions In The Six Quark System,''
  J.\ Phys.\ A {\bf 15}, 7 (1982).
   
\bibitem{BLL1}   S.~J.~Brodsky and C.~-R.~Ji,
  %``Quantum Chromodynamic Evolution of the Baryon System,''
  Phys.\ Rev.\ D {\bf 33}, 1951 (1986).
 
\bibitem{BLL2}   C.~-R.~Ji and S.~J.~Brodsky,
  %``Quantum Chromodynamic Evolution of Six Quark States,''
  Phys.\ Rev.\ D {\bf 34}, 1460 (1986).
  
 
\bibitem{Kusainov:1991vn} 
  A.~M.~Kusainov, V.~G.~Neudatchin and I.~T.~Obukhovsky,
  %``Projection of the six quark wave function onto the N N channel and the problem of the repulsive core in the N N interaction,''
  Phys.\ Rev.\ C {\bf 44}, 2343 (1991).
   
\bibitem{NSmass}
  P.~Demorest, T.~Pennucci, S.~Ransom, M.~Roberts and J.~Hessels,
  %``Shapiro Delay Measurement of A Two Solar Mass Neutron Star,''
  Nature {\bf 467}, 1081 (2010).

\bibitem{PH} H.~Heiselberg and V.~Pandharipande,
  %``Recent progress in neutron star theory,''
  Ann.\ Rev.\ Nucl.\ Part.\ Sci.\  {\bf 50}, 481 (2000).
 

\bibitem{isosrc}  E.~Piasetzky, M.~Sargsian, L.~Frankfurt, M.~Strikman and J.~W.~Watson,
  %``Evidence for the strong dominance of proton-neutron correlations in nuclei,''
  Phys.\ Rev.\ Lett.\  {\bf 97}, 162504 (2006).
  
\bibitem{eip4} 
  R.~Subedi, R.~Shneor, P.~Monaghan, B.~D.~Anderson, K.~Aniol, J.~Annand, J.~Arrington and H.~Benaoum {\it et al.},
  %``Probing Cold Dense Nuclear Matter,''
  Science {\bf 320}, 1476 (2008).
    

\bibitem{srcrev} 
  L.~Frankfurt, M.~Sargsian and M.~Strikman,
  %``Recent observation of short range nucleon correlations in nuclei and their implications for the structure of nuclei and neutron stars,''
  Int.\ J.\ Mod.\ Phys.\ A {\bf 23}, 2991 (2008).
  
  
   
\bibitem{FSEMCnew} L.~Frankfurt and M.~Strikman,
  %``QCD and QED dynamics in the EMC effect,''
  Int.\ J.\ Mod.\ Phys.\ E {\bf 21}, 1230002 (2012).
 

\bibitem{RP} J.~P.~Ralston and B.~Pire,
  %``Oscillatory Scale Breaking and the Chromo - Coulomb Phase Shift,''
  Phys.\ Rev.\ Lett.\  {\bf 49}, 1605 (1982).


\bibitem{BdTer}  S.~J.~Brodsky and G.~F.~de Teramond,
  %``Spin Correlations, QCD Color Transparency and Heavy Quark Thresholds in Proton Proton Scattering,''
  Phys.\ Rev.\ Lett.\  {\bf 60}, 1924 (1988).
 \bibitem{hepdata} HEPDATA, The Durham HEP database  http://durpdg.dur.ac.uk/hepdata/.
 \bibitem{Allaby} J.~V.~Allaby {\it et al.},  Phys.\ Lett.\  B {\bf 28}, 67 (1968).
\bibitem{Akerlof}C.~W.~Akerlof  {\it et al.}, Phys. Rev. {\bf 159}, 1138 (1967).
\bibitem{pnexp1} M.~L.~Perl {\it et al.},  Phys.\ Rev.\  D {\bf 1}, 1857 (1970).
\bibitem{pnexp2}J.~L.~Stone {\it et al.}, Nucl.\ Phys.\  B {\bf 143}, 1 (1978).

  \bibitem{ACM} 
  C.~Avilez, G.~Cocho and M.~Moreno,
  %``Polarized Nucleon-nucleon Scattering From 3-{GeV}/$c$ to 12-{GeV}/$c$: What Can Be Inferred From the Experiments?,''
  Phys.\ Rev.\ D {\bf 24}, 634 (1981).
  %%CITATION = PHRVA,D24,634;%%
  %32 citations counted in INSPIRE as of 31 Jan 2014

\bibitem{Lipkin} 
  H.~J.~Lipkin,
  %``Color Theory in a Spin,''
  Nature {\bf 324}, 14 (1986).

\bibitem{Bourrely} 
  C.~Bourrely and J.~Soffer,
  %``A New Mechanism for Spin Effects in Large Angle $N N$ Elastic Scattering,''
  Phys.\ Rev.\ D {\bf 35}, 145 (1987).
  
\bibitem{Preparata} 
  G.~Preparata and J.~Soffer,
  %``The Crucial Role of Spin in Large Angle Scattering,''
  Phys.\ Lett.\ B {\bf 180}, 281 (1986).
 
 
 \bibitem{Troshin} 
  S.~M.~Troshin and N.~E.~Tyurin,
  %``On Spin Spin Correlation Parameter Oscillations,''
  JETP Lett.\  {\bf 44}, 149 (1986).
   

\bibitem{AKP} 
  M.~Anselmino, P.~Kroll and B.~Pire,
  %``Diquarks in Exclusive Reactions at Large Momentum Transfer,''
  Z.\ Phys.\ C {\bf 36}, 89 (1987).


\bibitem{GR} 
  L.~E.~Gordon and G.~P.~Ramsey,
  %``Spin structure of the proton and large p(T) processes in polarized p p collisions,''
  Phys.\ Rev.\ D {\bf 59}, 074018 (1999).
        
\bibitem{CTB}S.~J.~Brodsky, in {\em Proceedings  of the Thirteenth International Symposium on Multiparticle 
Dynamics}, edited by W.~Kittel, W.~Metzeger and A. Stergiou, World Scientific, Singapore, p.963 (1981)

\bibitem{CTMul}A.~H.~Mueller, in {\em Proceedings of the Seventeenth Rencontre de Moriond}, Moriond 1982,
edited by J.~Tran Thanh Van, Editions Frontiers, Gif-sur-Yvette, France, p.13 (1982).


\bibitem{CTFMS}   L.~L.~Frankfurt, G.~A.~Miller and M.~Strikman,
  %``The Geometrical color optics of coherent high-energy processes,''
  Ann.\ Rev.\ Nucl.\ Part.\ Sci.\  {\bf 44}, 501 (1994).

\bibitem{CTRP}   P.~Jain, B.~Pire and J.~P.~Ralston,
  %``Quantum color transparency and nuclear filtering,''
  Phys.\ Rept.\  {\bf 271}, 67 (1996).


\bibitem{FLFS} 
  G.~R.~Farrar, H.~Liu, L.~L.~Frankfurt and M.~I.~Strikman,
  %``Transparency in Nuclear Quasiexclusive Processes with Large Momentum Transfer,''
  Phys.\ Rev.\ Lett.\  {\bf 61}, 686 (1988).

\bibitem{Kopeliovich}
  B.~Z.~Kopeliovich, J.~Nemchik and I.~Schmidt,
  %``Color Transparency at Low Energies: Predictions for JLAB,''
  Phys.\ Rev.\ C {\bf 76} (2007) 015205.

   
   
   
 
\bibitem{ct2jets} L.~Frankfurt, G.~A.~Miller and M.~Strikman,
  %``Coherent nuclear diffractive production of mini - jets: Illuminating color transparency,''
  Phys.\ Lett.\ B {\bf 304}, 1 (1993).
  
  \bibitem{ctrho}   L.~El Fassi, L.~Zana, K.~Hafidi, M.~Holtrop, B.~Mustapha, W.~K.~Brooks, H.~Hakobyan and X.~Zheng {\it et al.},
  %``Evidence for the Onset of Color Transparency in $\rho^0$ Electroproduction off Nuclei,''
  Phys.\ Lett.\ B {\bf 712}, 326 (2012).
  
\bibitem{ctpi}  B.~Clasie, X.~Qian, J.~Arrington, R.~Asaturyan, F.~Benmokhtar, W.~Boeglin, P.~Bosted and A.~Bruell {\it et al.},
  %``Measurement of nuclear transparency for the A(e, e-prime' pi+) reaction,''
  Phys.\ Rev.\ Lett.\  {\bf 99}, 242502 (2007).
  


\bibitem{Carroll} 
  A.~S.~Carroll, D.~S.~Barton, G.~Bunce, S.~Gushue, Y.~I.~Makdisi, S.~Heppelmann, H.~Courant and G.~Fang {\it et al.},
  %``Nuclear Transparency to Large Angle p p Elastic Scattering,''
  Phys.\ Rev.\ Lett.\  {\bf 61}, 1698 (1988).
   
\bibitem{Mardor} 
  I.~Mardor, S.~Durrant, J.~Aclander, J.~Alster, D.~Barton, G.~Bunce, A.~Carroll and N.~Christensen {\it et al.},
  %``Nuclear transparency in large momentum transfer quasielastic scattering,''
  Phys.\ Rev.\ Lett.\  {\bf 81}, 5085 (1998).
 
\bibitem{Aclander} 
  J.~Aclander, J.~Alster, G.~Asryan, Y.~Averiche, D.~S.~Barton, V.~Baturin, N.~Buktoyarova and G.~Bunce {\it et al.},
  %``Nuclear transparency in 90 degree c.m. quasielastic A(p,2p) reactions,''
  Phys.\ Rev.\ C {\bf 70}, 015208 (2004).
 
 \bibitem{FSZ} 
  L.~Frankfurt, M.~Strikman and M.~Zhalov,
  %``Single particle strength restoration and nuclear transparency in high Q**2 exclusive (e, e-prime p) reactions,''
  Phys.\ Lett.\ B {\bf 503}, 73 (2001).
     
 \bibitem{FMSS} 
  L.~L.~Frankfurt, E.~J.~Moniz, M.~M.~Sargsian and M.~I.~Strikman,
  %``Correlation effects in nuclear transparency,''
  Phys.\ Rev.\ C {\bf 51}, 3435 (1995). 
  
  
  
 \bibitem{Anselmino}
  M.~Anselmino, E.~Predazzi, S.~Ekelin, S.~Fredriksson and D.~B.~Lichtenberg,
  %``Diquarks,''
  Rev.\ Mod.\ Phys.\  {\bf 65}, 1199 (1993).
  
\bibitem{Wilczek} 
  F.~Wilczek,
  %``Diquarks as inspiration and as objects,''
  In *Shifman, M. (ed.) et al.: From fields to strings, vol. 1* 77-93   [hep-ph/0409168].
 
 \bibitem{Perl} Martin L. Perl, "High Energy Hadrons", John Wiley 
 \& Sons Inc , December 4, 1974, 582pp.
\bibitem{Bystricky}  
  J.~Bystricky, F.~Lehar and P.~Winternitz,
  %``Formalism of Nucleon-Nucleon Elastic Scattering Experiments,''
  J.\ Phys.\ (France) {\bf 39}, 1 (1978).
   
\bibitem{Bg1} N.~N.~Bogolyubov, V.~S.~Vladimorov and A.~N.~Tavkhelidze,
  %``On automodel asymptotics in quantum field theory. 1,''
  Teor.\ Mat.\ Fiz.\  {\bf 12}, 3 (1972).
\bibitem{Bg2}N.~N.~Bogolyubov, A.~N.~Tavkhelidze and V.~S.~Vladimirov,
  %``On automodel asymptotics in quantum field theory. 2,''  
  Teor.\ Mat.\ Fiz.\  {\bf 12}, 305 (1972).
\bibitem{DiracLF} P.~A.~M.~Dirac, 
%{\em Forms of Relativistic Dynamics}
Rev. Mod. Phys. {\bf 21}, 392 (1949).
\bibitem{SW}S.~Weinberg,
%{\em Dynamics at Infinite Momentum}
Phys. Rev. {\bf 150}, 1313 (1966).

\bibitem{BrodskyLF} S.J. Brodsky,   
  %``New Perspectives in quantum chromodynamics,''
  In *Beijing 1993, Proceedings, Particle physics at the Fermi scale* 271-374, and SLAC Stanford - SLAC-PUB-6304 (93/07,rec.Oct.) 104 p

\bibitem{BroPaul}   S.~J.~Brodsky, H.~-C.~Pauli and S.~S.~Pinsky,
  %``Quantum chromodynamics and other field theories on the light cone,''
  Phys.\ Rept.\  {\bf 301}, 299 (1998).

\bibitem{BLec2004}S.~J.~Brodsky,  
Lectures at the 58th Scottish University Summer School in Physics:
%A NATO Advanced  Study Institute and the  European Union Hadronic Physics 13th Summer Institute, 
St. Andrews, Schotland, 30 August --  1 September, 2004,   
hep-ph/0412101.
 
\bibitem{BroLep} 
  G.~P.~Lepage and S.~J.~Brodsky,
  %``Exclusive Processes in Perturbative Quantum Chromodynamics,''
  Phys.\ Rev.\ D {\bf 22}, 2157 (1980).
   
  
\bibitem{reducedAMPs} S.~J.~Brodsky and J.~R.~Hiller,
  %``Reduced Nuclear Amplitudes in Quantum Chromodynamics,''
  Phys.\ Rev.\ C {\bf 28}, 475 (1983)  [Erratum-ibid.\ C {\bf 30}, 412 (1984)].
   
\bibitem{h20}C.~White {\it et al.},  Phys.\ Rev.\  D {\bf 49}, 58 (1994).


\bibitem{gdpn} 
  L.~L.~Frankfurt, G.~A.~Miller, M.~M.~Sargsian and M.~I.~Strikman,
  %``QCD rescattering and high-energy two-body photodisintegration of the deuteron,''
  Phys.\ Rev.\ Lett.\  {\bf 84}, 3045 (2000).
 
\bibitem{gdpnpol} 
  M.~M.~Sargsian,
  %``Polarization observables in hard rescattering mechanism of deuteron photodisintegration,''
  Phys.\ Lett.\ B {\bf 587}, 41 (2004).
  
 
  \bibitem{gdpnpp} 
  S.~J.~Brodsky, L.~Frankfurt, R.~A.~Gilman, J.~R.~Hiller, G.~A.~Miller, E.~Piasetzky, M.~Sargsian and M.~Strikman,
  %``Hard photodisintegration of a proton pair in He-3,''
  Phys.\ Lett.\ B {\bf 578}, 69 (2004).
 
\bibitem{gdpn2} 
  M.~M.~Sargsian,
  %``Hard Break-Up of Two-Nucleons and QCD Dynamics of NN Interaction,''
  AIP Conf.\ Proc.\  {\bf 1056}, 287 (2008).
%\cite{Sargsian:2003sz}
 
\bibitem{hrm} 
  M.~M.~Sargsian and C.~Granados,
  %``Hard Break-Up of Two-Nucleons from the He-3 Nucleus,''
  Phys.\ Rev.\ C {\bf 80}, 014612 (2009).
  
 
\bibitem{pnangle} 
  C.~G.~Granados and M.~M.~Sargsian,
  %``Quark Structure of the Nucleon and Angular Asymmetry of Proton-Neutron Hard Elastic Scattering,''
  Phys.\ Rev.\ Lett.\  {\bf 103}, 212001 (2009).
 
\bibitem{gdbb} 
  C.~G.~Granados and M.~M.~Sargsian,
  %``Hard breakup of the deuteron into two $\Delta$-isobars,''
  Phys.\ Rev.\ C {\bf 83}, 054606 (2011).


\bibitem{RJ}R.~L.~Jaffe, Phys.\ Rept.\  {\bf 409}, 1 (2005).
\bibitem{SW2} A.~Selem and F.~Wilczek,  in the proceedings of 
"Ringberg Workshop on New Trends in HERA Physics"
2-7 Oct 2005, Ringberg Castle, Tegernsee, Germany,  arXiv:hep-ph/0602128.

\bibitem{FSDM}   L.~L.~Frankfurt, M.~I.~Strikman, D.~B.~Day and M.~Sargsian,
  %``Evidence for short range correlations from high Q**2 (e, e-prime) reactions,''
  Phys.\ Rev.\ C {\bf 48}, 2451 (1993).
 
 
\bibitem{srcprogress} 
  J.~Arrington, D.~W.~Higinbotham, G.~Rosner and M.~Sargsian,
  %``Hard probes of short-range nucleon-nucleon correlations,''
  Prog.\ Part.\ Nucl.\ Phys.\  {\bf 67}, 898 (2012).
 
\bibitem{edepn} 
  M.~M.~Sargsian,
  %``Large Q**2 Electrodisintegration of the Deuteron in Virtual Nucleon Approximation,''
  Phys.\ Rev.\ C {\bf 82}, 014612 (2010).
 
 \bibitem{edepnexp} 
  W.~U.~Boeglin {\it et al.}  [Hall A Collaboration],
  %``Probing the high momentum component of the deuteron at high Q^2,''
  Phys.\ Rev.\ Lett.\  {\bf 107}, 262501 (2011).
  
\bibitem{hnm} 
  M.~M.~Sargsian, J.~Arrington, W.~Bertozzi, W.~Boeglin, C.~E.~Carlson, D.~B.~Day, L.~L.~Frankfurt and K.~Egiyan {\it et al.},
  %``Hadrons in the nuclear medium,''
  J.\ Phys.\ G {\bf 29}, R1 (2003).
  
\bibitem{JLab12}White Paper: "The Science Driving the 12 GeV Upgrade of CEBAF", 
            Jefferson Lab, Newport News, VA, 2000.
  
    
\end{thebibliography}
\end{document}